\documentclass[lettersize,journal]{IEEEtran}
\usepackage{cite}
\usepackage{amsmath,amssymb,amsfonts}
\usepackage{algorithmic}
\usepackage{algorithm}
\usepackage{array}
\usepackage{graphicx}
\usepackage{textcomp}
\usepackage{xcolor}
\usepackage{float}
\usepackage{comment}
\usepackage{stfloats}
\usepackage{url}
\usepackage{verbatim}
\usepackage{enumerate}

\usepackage{enumitem}
\usepackage{multirow}
\usepackage{diagbox}
\usepackage{mathtools, cuted}
\usepackage{lipsum, color}
\usepackage{standalone}

\usepackage{subcaption}

\usepackage{tikz}
\usepackage{pgfplots}

\begin{document}
	\title{\LARGE \bf Predictor-Feedback CACC for Vehicular Platoons with Actuation and Communication Delays Based on a Multiple-Predecessor-Following CTH Nominal Strategy}
	
	\author{Amirhossein Samii$^1$, Dimitrios Angelopoulos$^1$, and Nikolaos Bekiaris-Liberis$^1$
		\thanks{Funded by the European Union (ERC, C-NORA, 101088147). Views and opinions expressed are however those of the authors only and do not necessarily reflect those of the European Union or the European Research Council. Neither the European Union nor the granting authority can be held responsible for them. 
			
			\vspace{1pt}
			$^1$A. Samii, D. Angelopoulos, and N. Bekiaris-Liberis are with the Department of Electrical \& Computer Engineering, Technical University of Crete, Chania, 73100, Greece. Email
			addresses: 
			{\tt\small asamii@tuc.gr}, {\tt\small dangelopoulos1@tuc.gr}, and {\tt\small bekiaris-liberis@ece.tuc.gr.}			}%
	}
	\markboth{SUBMITTED TO IEEE TRANSACTIONS ON INTELLIGENT TRANSPORTATION SYSTEMS ON APRIL 6 2026.}%
	{Shell \MakeLowercase{\textit{et al.}}: A Sample Article Using IEEEtran.cls for IEEE Journals}
	\IEEEpubid{}
    
	\maketitle
    
	\begin{abstract}
    We develop a predictor-feedback cooperative adaptive cruise control (CACC) design relying on a multiple-predecessor-following (MPF) topology-based, nominal, delay-free CACC law. We consider vehicular platoons with heterogeneous vehicles, whose dynamics are described by a third-order linear system subject to actuation delay, along with vehicle-to-vehicle
    (V2V) communication delay. The design achieves individual vehicle stability, string stability, and zero, steady-state speed/spacing tracking errors, for any value of the actuation delay. The proofs of individual vehicle stability, string stability, and regulation rely on employment of an input-output approach on the frequency domain, capitalizing on the delay-compensating property of the design, which enables as to derive explicit string stability conditions on control and vehicle models parameters. The theoretical guarantees of string stability and the respective conditions on parameters are illustrated also numerically. We present consistent simulation results, for a ten-vehicle platoon, illustrating the potential of the design in traffic throughput improvement, as compared with a predictor-feedback CACC design in which, each ego vehicle's controller utilizes information only from a single preceding vehicle. We also present simulation results in a realistic scenario in which the leading vehicle’s trajectory is obtained from NGSIM data. 
	\end{abstract}
    
	\section{Introduction}
    \subsection{Motivation}
Input delays in vehicular platoons equipped with Adaptive Cruise Control (ACC) and CACC systems may originate from several factors, including engine and sensor dynamics, brake actuation delays, and computational processing times, see, e.g., \cite{bekiaris_predictor_2018,dutch2,speed_error2,huang_ren_1998, Molnar Book, yanakiev_2001}; while communication delays originate in V2V communication, see, e.g., \cite{nominal_control_1, nominal_control_2, speed_error2, string_stability, Molnar Book}, \cite{Pan, italian, robust, dy2, speed_error3}. If left uncompensated, input and communication delays can compromise both individual vehicle stability and string stability, which are essential properties for maintaining safety and efficiency of vehicular platoons, see, e.g., \cite{bekiaris_predictor_2018,dehaan_observer_2023,molnar_safety_2023,oncu_string_2014,italian, robust,predictor_feedback3},\cite{predictor_feedback4},\cite{predictor_feedback5,CDC,xiao_gao_2011,dy1}. In particular, for long actuation delays (that may be the case in practice, particularly for internal combustion engine and heavy-duty vehicles, see, e.g., \cite{Molnar Book}, \cite{yanakiev_2001}), the use of a predictor-based control approach is necessary, as neither individual vehicle stability nor string stability can be guaranteed as the actuation delay value increases, irrespectively of the baseline ACC/CACC design employed. The MPF topology-based CACC design from \cite{string_stability} (see also, e.g., \cite{V2V}) in particular has been proven to be effective in featuring the potential of improving traffic throughput, since it may achieve, in general, string stability for smaller time headways \cite{nominal_control_1}, \cite{nominal_control_2}. Consequently, in this paper we construct a predictor-feedback CACC law using an MPF topology-based nominal CACC law, towards compensation of long actuation delays under the simultaneous presence of communication delays, while potentially achieving higher traffic throughput, as well as providing formal guarantees of both individual vehicle stability and string stability.

    \subsection{Literature on Delay-Compensating ACC/CACC Designs}
    Related literature includes predictor-based ACC/CACC designs that aim at compensation of long actuation delays in vehicular platoons, which are developed in \cite{bekiaris_predictor_2018, bekiaris_cth_2023, caiazzo_dos_2023,davis_2021, molnar_predictor_2018, molnar_safety_2023, Molnar Book, predictor_feedback3},\cite{ predictor_feedback4, predictor_feedback5}, \cite{CDC}, \cite{ wang_delay_2016, dy1, yanakiev_2001, speed_error3, zhao_yu_2024} utilizing the general, predictor-feedback design method (see, e.g., \cite{predictor_based_0}, \cite{b2}, \cite{Kristic}). Small actuation delays are addressed in \cite{dutch2, dehaan_observer_2023}, \cite{speed_error2, huang_ren_1998, xiao_gao_2011, zhang_switched_2023}, without requiring incorporation of a predictor structure in the design. Designs that account for communication delays only (but cannot achieve actuation delay compensation, which requires different treatment in design, as well as in stability and string stability analyses) are also relevant and they are developed, e.g., in \cite{nominal_control_1, nominal_control_2,Pan, string_stability, italian, dy2}; while both actuation and communication delays are addressed in \cite{robust, zhang_switched_2023}. Despite the existence of predictor-based ACC/CACC designs that aim at actuation delay compensation, to the best of our knowledge, none of the existing results rely on the MPF topology-based nominal CACC strategy from \cite{string_stability} (see also, \cite{nominal_control_1, nominal_control_2}), towards construction of a predictor-based counterpart that simultaneously achieves actuation delay compensation under the simultaneous presence of communication delays, while enabling higher traffic throughput.

    \subsection{Contributions}
    In the present paper, we construct a predictor-feedback CACC law under V2V communication delays that consists of two main elements, namely, a nominal CACC design for the (actuation/communication) delay-free case that relies on the MPF topology-based CACC law from \cite{string_stability} (see also, \cite{nominal_control_1, nominal_control_2}) and a predictor structure that enables to employ the predictors of all states incorporated in the nominal, delay-free design. The design requires measurements of the states from a certain number of vehicles ahead, as well as information about the delay/lag and desired time-headway from the same number of preceding vehicles. Apart from the delay/lag values, which are required for constructing the predictor states, the rest of the measurements and parameters information are requirements of the nominal, delay-free control law, which can be obtained via V2V or V2X communication (despite imposing communication delays), see, e.g., \cite{nominal_control_1, nominal_control_2, string_stability}. The control strategy developed achieves $\mathcal{L}_2$ string stability with respect to speed/acceleration errors propagation, stability of individual vehicles, and zero steady-state, spacing/speed errors, for a constant leader’s speed. The formal proofs rely on an input-output approach on the frequency domain, considering third-order vehicles' dynamics with actuation and communication delays, and capitalize on the delay-compensating property of the design. 
    
    In particular, we derive explicit conditions on control and model parameters, including the delay values, which we illustrate also numerically; while we provide a general guide for selecting the control parameters. We then provide simulation results for a platoon of ten vehicles, focusing on a practical scenario where a vehicle cuts in front of the platoon and then executes an acceleration/deceleration maneuver. We also compare performance of the present design with the predictor-based CACC design from \cite{bekiaris_cth_2023}, which cannot guarantee string stability for small time-headways, for a fixed set of control parameters. In addition, we validate the design considering a realistic scenario in which the leading vehicle’s acceleration is obtained from the NGSIM data. All case studies confirm the effectiveness of the design developed.

    \subsection{Organization}
    The outline of the paper is as follows. Section II presents the vehicular platoons considered and the predictor-feedback CACC design developed. In Section III, we state our main result, which is vehicle/string stability and regulation under the CACC law constructed, whose proof is provided in Appendix~A; as well as we present numerical verification of the theoretical, string stability conditions derived and a general guide for selecting the control parameters. Simulation results are presented in Section IV, including comparisons with \cite{bekiaris_cth_2023} and a study using NGSIM data. In Section V we provide concluding remarks.

    \section{CACC for Heterogeneous Platoons With Actuation And Communication Delays}
	\subsection{Vehicle Model, Available Measurements, and Nominal Control Design}
    \begin{figure}[htb!]
		\begin{center}
			\includegraphics[width = 9cm, height = 3.5cm]{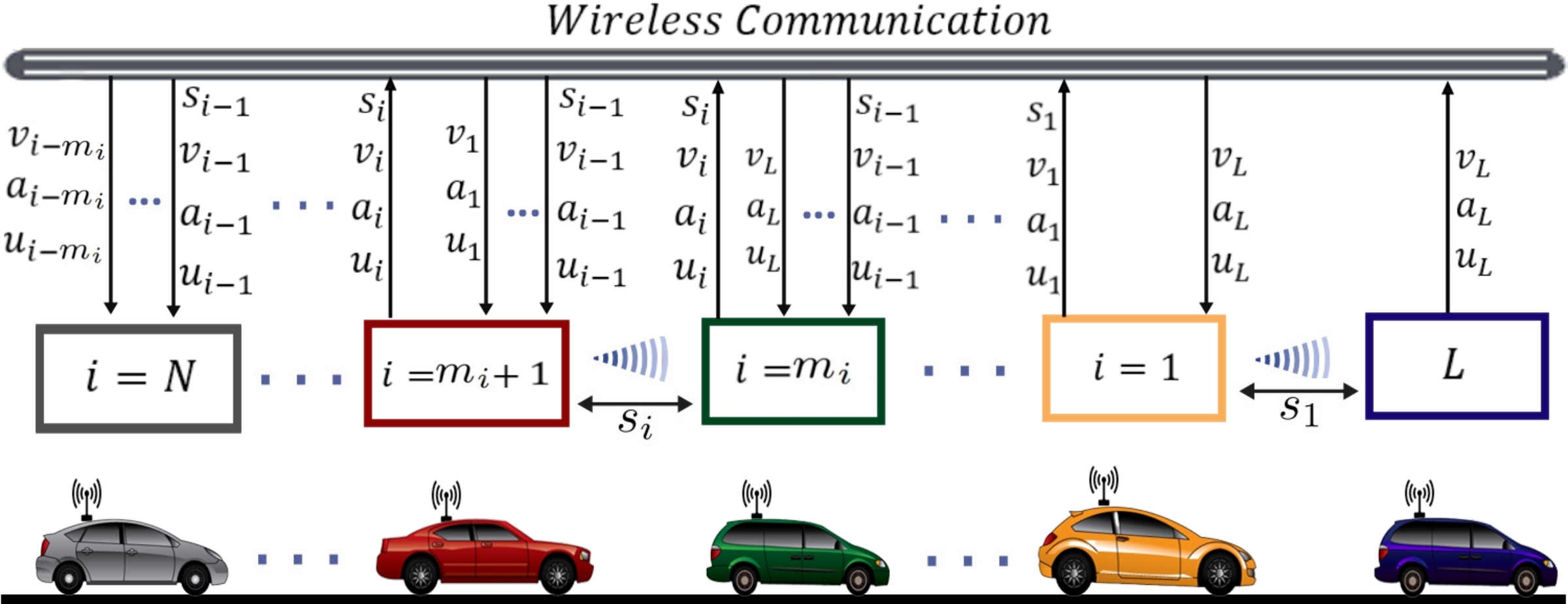}
			\caption{Platoon of $N+1$ heterogeneous vehicles following each other in a single lane without overtaking. The dynamics of each vehicle $i = 1,..., N$ are governed by system (\ref{dy1})--(\ref{dy3}). Each vehicle can measure its own speed/acceleration, the relative speed with the preceding vehicle, and the spacing with respect to the preceding vehicle. Moreover, each vehicle utilizes the control input, acceleration, spacing, and velocity information of $m_i$ vehicles ahead, obtained via V2V or V2X communication.}\label{Fig1}
		\end{center}
	\end{figure}
	\textit{a) Vehicle dynamics:} We consider a heterogeneous string of vehicles (see Fig. \ref{Fig1}) each one modeled by the following third-order, linear system with actuator delay that describes vehicle dynamics (see, e.g., \cite{nominal_control_1, dutch2,italian,predictor_feedback3,dy1})
	\begin{align}
		\dot{s}_i(t) =&\nobreakspace v_{i-1}(t) - v_i(t),\label{dy1}\\	
		\dot{v}_i(t) =&\nobreakspace a_i (t),\label{dy2}\\
		\dot{a}_i(t) =&\nobreakspace -\frac{1} {\tau_i} a_i (t)+\frac{1} {\tau_i}u_i(t-D),\label{dy3}
	\end{align}
    	
	\noindent$i = 1,...,N$, where $s_i=x_{i-1} - x_i - l_i$ and $x_i$ is the position of vehicle $i$ and $l_i$ is its length, $v_i$ is vehicle speed, $a_i$ is vehicle acceleration, $\tau_i$ is lag, capturing, engine dynamics, $u_i$ is the individual vehicle’s control variable, $D \geq 0$ is input delay, and $t \geq 0$ is time. Note that for the leading vehicle we assume similarly that it has the same type of third-order dynamics as the rest of the vehicles\footnote{The design can be modified in a straightforward manner when this is not true.}. We adopt the convention that $v_0 = v_L$ and $a_0 = a_L$ are the speed and acceleration of the string leader, respectively.

	\textit{b) Available measurements/information:} For the platoon considered here the measurements available to the ego vehicle $i$ are its own spacing $s_i$, speed $v_i$, acceleration $a_i$, and control input $u_i$, as well as the speed of the preceding vehicle $v_{i-1}$. It is possible to obtain this information through on-board sensors. Furthermore, control inputs of $m_i$ preceding vehicles, as well as their own acceleration, spacing, and speed are also available to vehicle $i$, together with information about their lags and desired time-headways. These measurements (and parameters information) are transmitted from the preceding vehicles, through V2V or V2X communication (see, e.g., \cite{nominal_control_1, nominal_control_2, string_stability, Molnar Book}). Moreover, the transmitted information, $u_{i-n_i}$, $s_{i-n_i}$, $v_{i-n_i}$ and $a_{i-n_i}$, $n_i \in \{1,...,m_i\}$ is obtained from the preceding vehicles via V2V communication, which may be subject to communication delays. Thus, the actual measurements are $u_{i-n_i,{\rm m}}$, $s_{i-n_i,{\rm m}}$, $v_{i-n_i,{\rm m}}$ and $a_{i-n_i,{\rm m}}$, which are defined as  $a_{i-n_i,{\rm m}}(t) = a_{i-n_i}(t-D_{{\rm c},i-n_i})$, $s_{i-n_i,{\rm m}}(t) = s_{i-n_i}(t-D_{{\rm c},i-n_i})$, $v_{i-n_i,{\rm m}}(t) = v_{i-n_i}(t-D_{{\rm c},i-n_i})$, and $u_{i-n_i,{\rm m}}(\theta) =~ u_{i-n_i}(\theta-~D_{{\rm c},i-n_i}), \theta \in [t-D,t]$, respectively, where $D_{{\rm c},i-n_i} \geq 0$, for $n_i \in \{1,...,m_i\}$, $i=1,...,N$, are communication delays.

    \textit{c) Nominal control design:} Without input and communication delays, the following MPF topology-based control strategy is constructed (see, e.g., \cite {nominal_control_1}\footnote{Representation (\ref{CLN}) is equivalent to (7) in \cite{nominal_control_1}, which follows from tedious algebraic manipulations and by setting the vehicles' length $l_i$ equal to the standstill gap $d_i$ in \cite{nominal_control_1}. Controller (\ref{CLN}) can be trivially modified to also incorporate a standstill gap.},\cite{nominal_control_2})
		\begin{align}
			u_i(t)&= \tau_i \alpha_i
\sum_{n_i=1}^{m_i} (m_i-n_i+1)\,\frac{h_{i-n_i+1}}{h_i}\left(
\frac{s_{i-n_i+1}(t)}{h_{i-n_i+1}}\right. \nonumber\\
&\left.- v_{i-n_i+1}(t)
\right)+ \tau_i b_i
\left(
\sum_{n_i=1}^{m_i} v_{i-n_i}(t) - m_i v_i(t)
\right)
\nonumber\\
&+ \tau_i c_i
\left(
\sum_{n_i=1}^{m_i} a_{i-n_i}(t) - m_i a_i(t)
\right),
			\label{CLN}
		\end{align}

\noindent for $i\geq m_i$, where $\alpha_i>0$, $b_i>0$, and $c_i>0$ are design parameters, and $h_i > 0$ is a desired time-headway. Controller~(\ref{CLN}) aims to regulate $v_i$ to the average speed of the preceding vehicles with which the ego vehicle communicates. While the first term in (\ref{CLN}) indicates the objective of regulating the spacing of each vehicle to $h_iv_i$, as customary in constant time-headway (CTH)-based strategies. In principle, by receiving information from multiple vehicles ahead, the ego vehicle can react proactively to the maneuvers of preceding vehicles. This may be beneficial for string stability of the platoon, as compared to reacting only to a single predecessor, in a case that the ego vehicle receives information only from one single vehicle ahead. This is illustrated in Sections III--V.

\subsection{Predictor-Feedback CACC Design}
    \begin{figure}[H]
    	\centering
    \vspace*{-\baselineskip}
			\includegraphics[width = 9cm, height = 6cm]{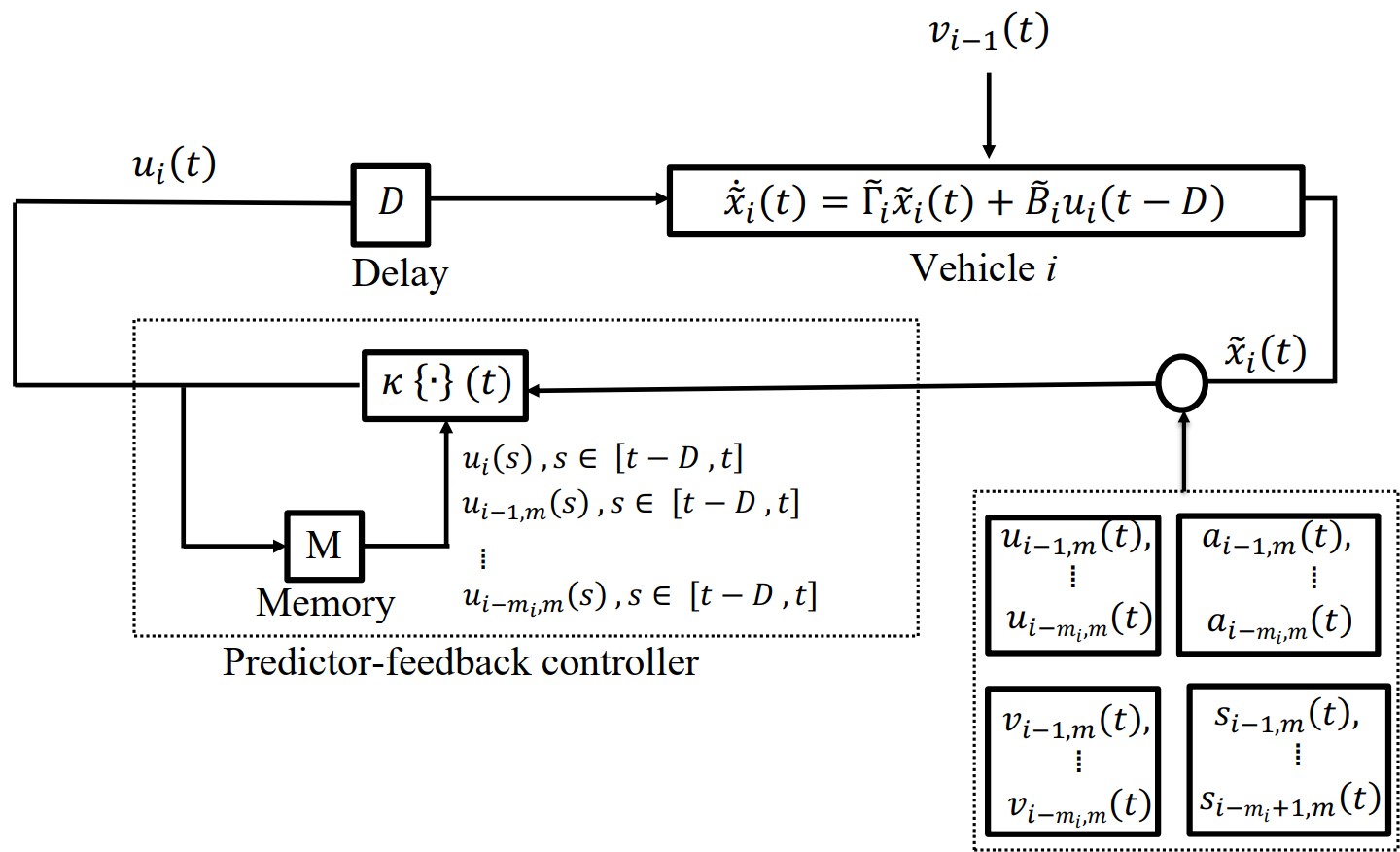}
			\caption{Block diagram of the predictor-feedback control design. The operator $\kappa\{\cdot\}$ defines the predictor-feedback law of each ego vehicle that utilizes the available past actuation commands of the ego vehicle and preceding vehicles. Moreover, the right dashed box illustrates all the needed information from the preceding vehicles, which is obtained via onboard sensors and V2V or V2X communication systems. Here we denote $\tilde{x}_{i}=\protect\begin{bmatrix} s_{i} \\ v_{i} \\ a_{i} \protect\end{bmatrix},\;
				\tilde{\Gamma}_i=\protect\begin{bmatrix} 
					0 & -1 & 0 \\
					0 & 0 & 1 \\
					0 & 0 & -\frac{1}{\tau_i}
				\protect\end{bmatrix},\;
				\tilde{B}_{i}=\protect\begin{bmatrix} 
					0 \\ 0 \\ \frac{1}{\tau_i}
				\protect\end{bmatrix}$.}\label{Block}
	\end{figure}
The predictor-feedback control laws (see also Fig.~\ref{Block}) for system (\ref{dy1})--(\ref{dy3}) are given by
\begin{align}
u_i(t)= \nobreakspace K_i^{\rm T}q_i(t),\label{CL}
\end{align}
where
\begin{align}
q_i(t)
&= e^{\Gamma_iD}\,\bar{x}_i(t)
+\int_{t-D}^{t}
e^{\Gamma_i(t-\theta)}\,
B_{i}\,
u_{i}(\theta)\, d\theta \nonumber \\
&+ \sum_{j=1}^{m_i}
\int_{t-D}^{t}
e^{\Gamma_i(t-\theta)}\,
B_{i-j}\,
u_{i-j,{\rm m}}(\theta)\, d\theta,
\label{PF_m}
\end{align}
with 
\newcommand\bigzero{\makebox(3,3){\text{\huge0}}}
\newcommand{\rvline}{\hspace*{-\arraycolsep}\vline\hspace*{-\arraycolsep}}
    \begin{align}
		&q_{i}=\begin{bmatrix} 
			q_{i,{s_i} }  \\
            q_{i,s_{i-1} }  \\ 
			\mathord{\vdots}\\
			q_{i,s_{i-m_i+1}}\\
			q_{i,v_{i}} \\
            q_{i,v_{i-1}} \\
			\mathord{\vdots}\\
			q_{i,v_{i-m_i}}\\
			q_{i,a_{i}} \\
            q_{i,a_{i-1}} \\
			\mathord{\vdots}\\
			q_{i,a_{i-m_i}}
		\end{bmatrix},\quad
		\bar{x}_{i}=\begin{bmatrix} 
			s_{i}\\
            s_{i-1,{\rm m}}\\
			\mathord{\vdots}\\
			s_{i-m_i+1,{\rm m}}\\
			v_{i}\\
            v_{i-1,{\rm m}}\\
			\mathord{\vdots}\\
			v_{i-m_i,{\rm m}}\\
			a_{i}\\
            a_{i-1,{\rm m}}\\
			\mathord{\vdots}\\
			a_{i-m_i,{\rm m}}
		\end{bmatrix},\label{xbar}\\
		&\Gamma_i=\nonumber\\
		&\begin{bmatrix} 
			\mathbf{0}_{m_i\times m_i}& \rvline & M_{{i,1}_{m_i\times(m_i+1)}} & \rvline &\mathbf{0}_{m_i\times(m_i+1)} \\
			\hline
			\mathbf{0}_{(m_i+1)\times m_i}& \rvline &\mathbf{0}_{(m_i+1)\times(m_i+1)} & \rvline & M_{{i,2}_{(m_i+1)\times(m_i+1)}}\\
			\hline
			\mathbf{0}_{(m_i+1)\times m_i} & \rvline& \mathbf{0}_{(m_i+1)\times(m_i+1)} & \rvline & M_{{i,3}_{(m_i+1)\times(m_i+1)}}\\
		\end{bmatrix},
    \end{align}
	and
	\begin{align}
		M_{{i,1}_{m_i\times(m_i+1)}}=&\begin{bmatrix}-1 & 1 & 0 &... & 0 \\
			&\ddots&\ddots&\ddots\\
			0 & 0 & ... & -1 & 1\end{bmatrix},\\
        M_{{i,2}_{(m_i+1)\times(m_i+1)}}=&\mathbb{I}_{(m_i+1)\times(m_i+1)},\\
        M_{{i,3}_{(m_i+1)\times(m_i+1)}}=&\begin{bmatrix}-\frac{1}{\tau_{i}}& 0 &...& 0\\
			&\ddots&\ddots\\
			0 & ... & 0 &  -\frac{1}{\tau_{i-m_i}}\end{bmatrix},\\
		B_{i-j}=& \begin{bmatrix}\vspace{10px}
			\hspace{1.5 mm} \mathbf{0}_{2m_i+1+j}\\
			{\frac{1}{\tau_{i-j}}}\\
			\mathbf{0}_{m_i-j}\vspace{8px}
		\end{bmatrix},\quad B_{i}= \begin{bmatrix} 
			\vspace{10px}
			\hspace{1.5 mm}\mathbf{0}_{2m_i+1}\\
			{\frac{1}{\tau_{i}}}\\
			\\
			\mathbf{0}_{m_i}\vspace{8px}
		\end{bmatrix},\label{B1}
        \end{align}
        \begin{align}
		K_{i}=&\begin{bmatrix}
			\vspace{10px}
			\hspace{1.5 mm}m_i\frac{\tau_i\alpha_i}{h_i}\\
			(m_i-1)\frac{\tau_i\alpha_i}{h_i} \\
			\mathord{\vdots}\\
			\frac{\tau_i\alpha_i}{h_i}\\
			-m_i\tau_{i}(\alpha_{i}+b_{i}) \\
			\tau_{i}b_i-\tau_i\alpha_i(m_i-1)\frac{h_{i-1}}{h_i} \\
            \mathord{\vdots}\\
            \tau_{i}b_i\\
			-m_i\tau_ic_{i} \\
			\tau_{i}c_{i}\\
			\mathord{\vdots}\\
            \tau_{i}c_{i}
		\end{bmatrix}.\label{K1}
	\end{align}

The structure of the predictor-feedback laws (\ref{CL})--(\ref{K1}) is explained as follows. In the absence of communication delays, each ego vehicle's control input would employ the predictor states of all states involved in the nominal design (\ref{CLN}), which would constitute vector $\bar{x}_i$ that would satisfy
\begin{align}
\dot{\bar {x}}_i(t)=\Gamma_i \bar{x}_i(t)+\bar{B}_i \bar{u}_i(t-D), \label{eq:cascade}
\end{align}
where 
\begin{align}
\bar{B}_i&=[B_i\ldots B_{i-m_i}],\\
\bar{u}_i&=[u_i\ldots u_{i-m_i}]^{\rm T}.
\end{align}
By then considering the dynamics of $\bar{x}_i$, because the delay is identical in each input channel, one could construct a predictor-feedback law as in the case of single-input systems (see, e.g., \cite{b2}), which would result in $q_i(t)=\bar{x}_i(t+D)$. However, since in the present case delays affect all measurements stemming from V2V communication, one has to modify this nominal, predictor-feedback CACC design as in (\ref{PF_m}), (\ref{xbar}). Furthermore, in the absence of actuation delay, control laws (\ref{CL}) would not be identical to [1, (8)] because we intentionally introduce communication delay to measurements $s_{i-1}$ and $v_{i-1}$ (that are available from on-board sensors). We make this modification here because it turns out to be beneficial for string stability. In particular, with this modification we obtain simpler-to-verify, explicit string stability conditions, even for the case of heterogeneous communication delays.


\section{String Stability Analysis}
\subsection{Main Results}
We start providing the definition of string stability employed under MPF topology. A platoon of vehicles indexed by $i=1,...,N,$ following each other within one lane without overtaking, is strictly $\mathcal{L}_2$ string stable with reference to speed errors if the following condition holds 
	\begin{align}
		\sum_{n_i=1}^{m_i} \left\lVert G_{i,i-n_i}\right\rVert_{\infty} \le 1,\label{G}
	\end{align}
where $\|G_{i,i-n_i}\|_{\infty} = \sup_{\omega\geq0} |G_{i,i-n_i}(j\omega)|$, while $G_{i,i-n_i}(j\omega)$ denotes the transfer function between the $i$-th vehicle's speed and the speed of its preceding vehicle $i-n_i$. Definition (\ref{G}) of string stability is the counterpart of the definition in \cite{nominal_control_1, nominal_control_2, string_stability}, which consider string stability with respect to spacing errors propagation. For obtaining simple, nevertheless still relevant conditions for string stability, within the scope of studying traffic flow performance, we define string stability with respect to speed errors propagation, as it is also the case in, e.g., \cite{speed_error1, speed_error2, dutch, L_stability, van Arem, speed_error3}. We note that for homogeneous platoons the two definitions are identical (see, e.g., \cite{nominal_control_1}). 

\textit{Theorem 1:} Consider a platoon of vehicles with heterogeneous dynamics modeled by (\ref{dy1})--(\ref{dy3}), under control laws (\ref{CL}) with (\ref{PF_m})--(\ref{K1}). For any $D \geq 0$, the platoon is $\mathcal{L}_2$ string stable with respect to speed errors propagation provided that the following conditions hold for $i=1,\ldots,N$
 \begin{align}
          \left(\frac{1}{\tau_i}+m_ic_i\right)(\alpha_i+b_i)-\frac{\alpha_i}{h_i} >0,  \label{stability2}
        \end{align}
along with 
\begin{align}
\bar{\beta}_i \geq0,\quad\beta_i \geq0\quad \text{and}\quad \bar{\gamma}_{i} \geq0,\quad \gamma_{i,n_i} \geq0,\nonumber\\
\forall \ n_i \in \{2,...,m_i\}, \label{c_5}
\end{align}
or
\begin{align}
	4\bar{\gamma}_{i}-\bar{\beta}_i^2\geq0, \quad 4\gamma_{i,n_i}-\beta_i^2\geq0 \quad \text{and}\nonumber\\
    \bar{\beta}_i <0,\quad \beta_i <0,\quad  \forall \ n_i \in \{2,...,m_i\},\label{c_3} 
\end{align}
where
\begin{align}
    \bar{\beta}_i&=\left(\frac{1+m_i\tau_i c_i}{\tau_i}\right)^2-2m_i(\alpha_i+b_i)-m_i^2 c_i^2\nonumber\\
    &-m_i^2 2 c_i \frac{m_i \alpha_i}{h_i} D_{{\rm c},i-1}(2D_{{\rm c},i-1}+D),\label{c_0}\\
	\beta_i &= \frac{1}{\tau_i^2} + 2m_i\frac{c_i}{\tau_i}-2m_i(\alpha_i+b_i),\label{c_1}\\
    \bar{\gamma}_{i}&=m_i^2(\alpha_i+b_i)^2
-2\left(\frac{1+m_i\tau_i c_i}{\tau_i}\right)
\frac{m_i\alpha_i}{h_i}\nonumber\\
&-m_i^2\Bigg(
-2 c_i \frac{\alpha_i}{h_i}
+ \left( b_i - (m_i-1)\alpha_i \frac{h_{i-1}}{h_i} \right)^2
\nonumber\\
&+ 2\left( \frac{m_i \alpha_i}{h_i} D_{{\rm c},i-1} \right)^2
+ 4 \left| b_i - (m_i-1)\alpha_i \frac{h_{i-1}}{h_i} \right|\nonumber\\
&\times \frac{m_i \alpha_i}{h_i} D_{{\rm c},i-1} \Bigg)\nonumber\\
&-2m_i^2\frac{\alpha_i}{h_i}
    \left( \frac{m_i \alpha_i}{h_i} D_{{\rm c},i-1} \right)(2D_{{\rm c},i-1}+D),\label{c_02}\\
    \gamma_{i,n_i}&= - 2m_i\frac{\alpha_i}{h_i\tau_i} + 2m_i^2\Big(1+(m_i-n_i)\frac{h_{i-n_i}}{h_i}\Big)\alpha_ib_i \nonumber\\
    &+m_i^2 \left( 1 - (m_i-n_i)^2 \frac{h_{i-n_i}^2}{h_i^2}\right)\alpha_i^2.\label{c_2}
\end{align}	
Furthermore, for a constant leading vehicle’s speed, say $v_{\rm ss}$, regulation is achieved with $\lim_{t\to+\infty} a_i(t)=0$, $\lim_{t\to\infty} v_i(t)=v_{\rm ss}$, and $\lim_{t\to+\infty} s_i(t)= h_iv_{\rm ss}$, for $i=1,\dots,N$.
\begin{center} \textit{Proof:} The proof can be found in Appendix A. \end{center}

An immediate consequence of Theorem 1 is formulated in the next corollary.

\textit{Corollary 1:} Consider a platoon of vehicles with heterogeneous dynamics modeled by (\ref{dy1})--(\ref{dy3}), under control laws (\ref{CL}) with (\ref{PF_m})--(\ref{K1}). If $D_{c,i}=0$, $i=0,...,N$, then for any $D \geq 0$ and $h_i>0$, the platoon is $\mathcal{L}_2$ string stable with respect to speed errors propagation provided that (\ref{stability2}) together with the following conditions hold for $i=1,\ldots,N$
\begin{align}
\beta_i \geq0\quad \text{and}\quad \gamma_{i,n_i} \geq0,\quad
\forall \ n_i \in \{1,...,m_i\}, \label{c_5c}
\end{align}
or
\begin{align}
	4\gamma_{i,n_i}-\beta_i^2\geq0 \quad \text{and}\quad \beta_i <0,\quad  \forall \ n_i \in \{1,...,m_i\}.\label{c_3c} 
\end{align}

\textit{Remark 1:} Corollary 1 provides sufficient conditions for stability in the special case where there are no communication delays. In this special case, condition (\ref{c_5}) or (\ref{c_3}) can be more easily verified. In particular, this can be seen from the fact that a necessary condition for (\ref{c_5c}) (or (\ref{c_3c})) to be satisfied can be derived as $h_i\geq  \frac{2\tau_i}{1+2\tau_im_ic_i}$, which illustrates that for fixed control gain $c_i$, the allowable time headway $h_i$ can be reduced by increasing the number $m_i$. This in turn may be beneficial which respect to traffic throughput improvement. We note however that there is no restriction on the desirable  $h_i$, since it can be made arbitrarily small by increasing $c_i$ (while still satisfying (\ref{stability2})). In fact, conditions (\ref{stability2}), (\ref{c_5c}) can be satisfied, for example, with a sufficiently large choice of $b_i$ and $c_i$. The situation is more complex in the presence of communication delays. In particular, the feasibility of simultaneously satisfying the conditions in Theorem~1 for given values of $h_i$ and $D_{c,i}$ can be established as follows. By selecting sufficiently small values of $\alpha_i$, the terms associated with actuation and communication delays in~(\ref{c_0}) and~(\ref{c_02}) can be effectively mitigated. Moreover, since $h_i$ appears in the conditions multiplied by $\alpha_i$, the influence of the time headway can also be reduced. Thus, for example, for satisfying (\ref{stability2}), (\ref{c_5}) one can select a sufficiently large $c_i$ and a sufficiently large $b_i$. However, when $m_i$ is too large, and in order for the actuation delay length to not be restricted, this may necessarily restrict the ratio $m_i\frac{D_{{\rm c},i-1}}{h_i}$ (see also \cite{nominal_control_1}) appearing in (\ref{c_0}), (\ref{c_02}). However, the actuation delay length is not restricted. These observations are consistent with the respective numerical analysis presented next.
   
\subsection{Numerical Illustration of String Stability}
\begin{figure}[htb!]
		\begin{center}
				\includegraphics[width = 9cm, height = 6cm]{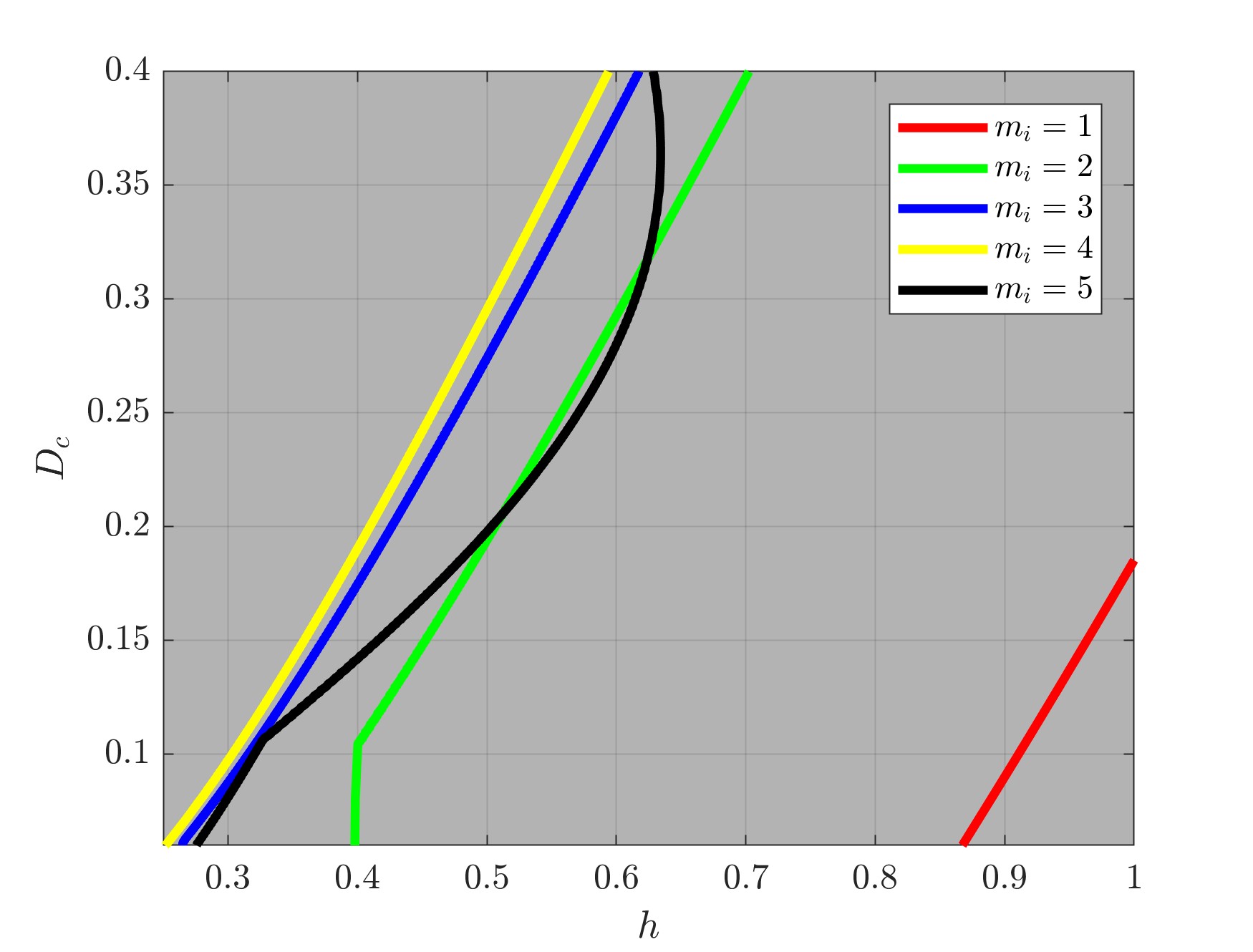}
                \includegraphics[width = 9cm, height = 6cm]{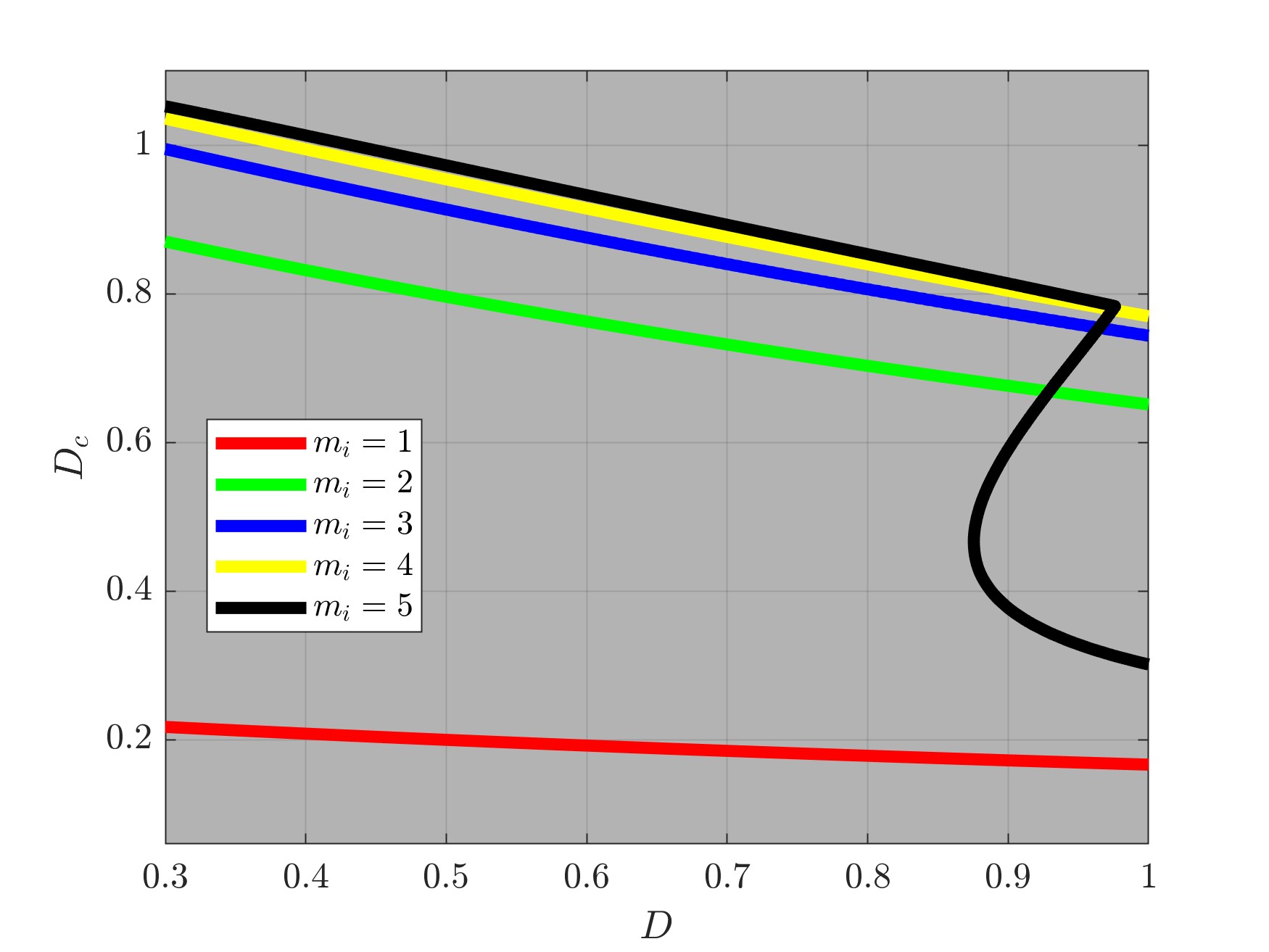}
				\caption{The region to the right of the colored lines indicates where string stability is satisfied for $m_i\in \{1,2,3,4,5\}$ and $D=0.7$ as function of $h$ and $D_{{\rm c}}$ (top); the region below of the colored lines indicates where string stability holds for $m_i\in \{1,2,3,4,5\}$  for $h=1$ as function of $D$ and $D_{{\rm c}}$ (bottom).} \label{s_stability3}

			\end{center}
	\end{figure}

A numerical analysis of the $\mathcal{L}_2$ string stability properties of the closed-loop systems, as actuation/communication delays and time headway vary, is provided next. In Fig. \ref{s_stability3} we depict the string stability regions, computed such that (\ref{A.17}) is satisfied (that may lead to less conservative conditions than (\ref{stability2})--(\ref{c_3})), for $m_i \in \{1,2,3,4,5\}$, where vehicles have identical engine lags $\tau_i=0.1$ and control gains $\alpha_i=5$, $b_i=10$, $c_i=2$. The first scenario (top) considers $D=0.7$ and illustrates the string stability region for different, over a certain realistic range, values of $h_i = h$ and $D_{{\rm c},i} \equiv D_{{\rm c}}$. We observe that as communication delay increases the allowable time-headway of every vehicle is increased as well, which is reasonably expected (see also, e.g., \cite{nominal_control_1, robust}). Moreover, we observe that as $m_i$ increases, for a given value of communication delay, the allowable time headway can be reduced (note that in the numerical scenarios considered the platoon is string stable according to (\ref{G}) for $m_i \le i \le N$). Nevertheless, as $m_i$ increases further (in the scenario depicted in Fig. \ref{s_stability3} for $m_i \geq 5$), for a given communication delay value the allowable time headway has to be increased. This is practically consistent as for a very large value of $m_i$ a vehicle communicates with too many vehicles via a delay, which may deteriorate the string stability properties of the platoon. The second scenario (bottom) considers $h_i = 1$ for all $i$ and illustrates the string stability region for different values of $D$ and $D_{{\rm c},i} \equiv D_{{\rm c}}$. It is observed that, in general, as actuation delay increases, for a given $m_i \le 4$, the maximum allowable communication delay decreases in order to maintain string stability. Furthermore, for $m_i \le 4$ and for a given value of actuation delay, as $m_i$ increases a larger value of communication delay is allowed. However, if $m_i \geq 5$, then the allowable communication delays may have to be reduced.

\subsection{Choice of Control Parameters}
\begin{figure}[H]
		\vspace*{-\baselineskip}\begin{center}
			\includegraphics[width = 9cm, height = 6cm]{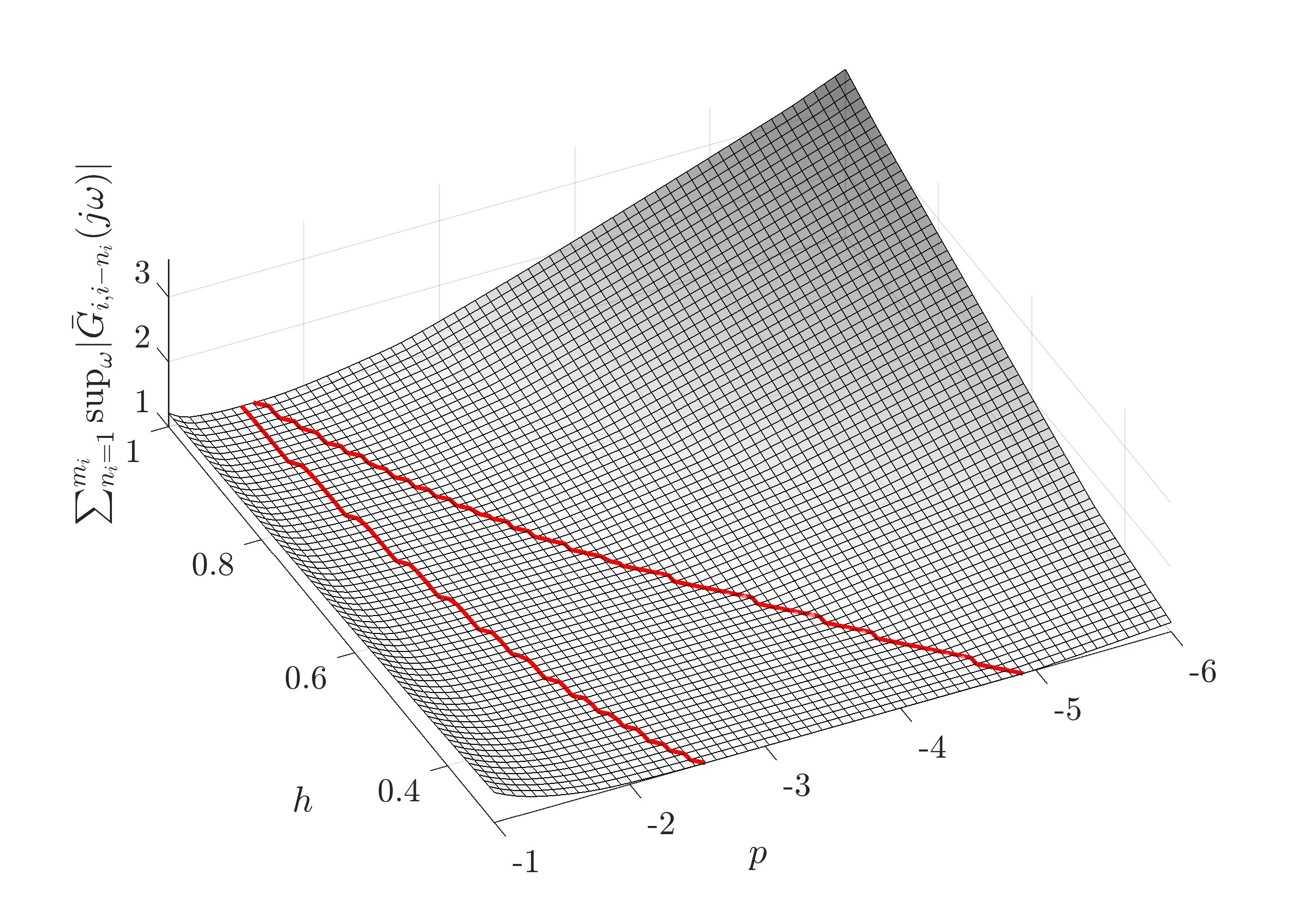}
			\caption{The values of $\sum_{n_i=1}^{m_i}\| \bar{G}_{i,i-n_i}\|_{\infty}$ for $m_i=3$ corresponding to (\ref{G_p}), for different values of $h$ and $p$.}\label{analitic}
		\end{center}\vspace*{-\baselineskip}
	\end{figure}

As a more constructive approach to choose $\alpha_i$, $b_i$, $c_i$, we recommend the following. We adopt the following parameterization of the gains for simplicity
	\begin{align}
		\alpha_i= -\frac{h_i}{m_i}p_i^3,\ b_i= \frac{h_i}{m_i}p_i^3+\frac{3}{m_i}p_i^2,\ c_i= -\frac{1}{m_i\tau_i} - \frac{3}{m_i}p_i\label{v3},
	\end{align}
	for some $p_i<0$ and all $i$, which results in the following transfer functions
    \begin{align}
    	\bar{G}_{i,i-n_i}(s)
    &=\frac{ \bar\mu_{1,i}(s)s^2
    + \bar\mu_{2,i}(s)s
    + \bar\mu_{3,i}(s) }{(s-p_i)^3},\label{G_p}
    \end{align}
    where
    \begin{align}
    	\bar\mu_{1,i}(s)&=-\frac{1}{m_i}\left(\frac{1}{\tau_i} + 3p_i\right)e^{-sD_{{\rm c},i-1}},\\
        \bar\mu_{2,i}(s)&=\frac{p_i^2}{m_i}\left(3+h_ip_i+(m_i-n_i)p_i\right)e^{-sD_{{\rm c},i-1}},
        \end{align}
        \begin{align}
        \bar\mu_{3,i}(s)&=
        \begin{cases}
        \displaystyle -\frac{p_i^3}{m_i}\left( e^{-sD_{{\rm c},i-1}} + m_i e^{-sD}\bigl(1-e^{-sD_{{\rm c},i-1}}\bigr)\right),\\
        \displaystyle \quad\quad\quad\quad\quad\quad\quad\quad\quad\quad\quad\quad\quad\quad\quad\quad\quad n_i=1,\\[6pt]
        \displaystyle -\frac{p_i^3}{m_i}e^{-sD_{{\rm c},i-n_i}},
        \displaystyle \quad\,\quad\quad\quad\quad\quad\quad\quad\quad\quad n_i\ge2.
        \end{cases}
    \end{align}
    Fig. \ref{analitic} depicts the set of parameter values for which both stability and string stability are satisfied, which is the region enclosed by the two red curves, for $m_i=3$ with $\tau_i=0.2$, control gains given by (\ref{v3}), communication delays $D_{{\rm c},i}=0.1$, and for fixed actuation delay $D=0.5$  for all $i$. As a general guideline on the choice of $p$ we observe the following. When $|p|$ takes relatively small values within this bounded region, the resulting controller gains are also small. Under these conditions, the sensitivity to measurement noise is mitigated, and a broader interval of admissible time headway values $h$ can be chosen while still maintaining stability and string stability. On the other hand, selecting a larger magnitude of $|p|$ leads to a quicker response of the vehicles to the leader’s actions and enhances disturbance rejection capabilities. Such higher values of $|p|$ are particularly appropriate when smaller desired time headway values $h$ are considered.

\section{Simulation Results}\label{sim}
	In this section, the performance of the predictor-feedback CACC design (\ref{CL}) with $m_1=1$, $m_2=2$, and $m_i=3$, $i\in \{3,4,5,6,7,8,9\}$ is demonstrated, followed by a comparison with the case $m_i=1$ for all $i$. Simulations are conducted in MATLAB R2022b, where integrals in the predictor-feedback controller (\ref{CL}) are computed using the trapezoidal rule. Moreover, third-order models for vehicles’ dynamics, as described in (\ref{dy1})–(\ref{dy3}), are implemented using the Euler method. Additionally, a fixed time step of $T_s = 0.01$ is selected to align with realistic sampling times of control execution and measurements, see, e.g., \cite{dutch2}, \cite{CDC}. 
    
    \begin{table}[htb!]
	\centering
	\caption{Parameters used for the simulation results in Fig.~\ref{Fig7}.}
	\scalebox{1}{\begin{tabular}{||c| c c c c||} 
			\hline
			\backslashbox{Vehicle No.}{Parameters} & $\tau_{i}$ & $h_i$ & $D_{\rm c,i}$ & $m_i$ \\ 
			\hline\hline
			0&$0.3$ & $-$ & $0.03$ & $-$\\ 
			\hline
			1&$0.3$ s & $0.4$ s & $0.09$ & $1$\\ 
			\hline
			2&$0.25$ s & $0.4$ s & $0.12$ & $2$\\
			\hline
			3&$0.25$ s & $0.5$ s & $0.14$ & $3$\\
			\hline
			4&$0.2$ s & $0.5$ s & $0.09$ & $3$\\
			\hline
			5&$0.25$ s & $0.3$ s & $0.18$ &$3$\\
			\hline
            6&$0.3$ s & $0.25$ s & $0.1$ &$3$\\
			\hline
            7&$0.25$ s & $0.25$ s & $0.12$ &$3$\\
			\hline
            8&$0.25$ s & $0.5$ s & $0.14$ &$3$\\
			\hline
            9&$0.3$ s & $0.3$ s & $-$ &$3$\\
			\hline
	\end{tabular}}
	\label{table2}
\end{table}

\begin{figure}[p]
	\begin{center}
		\includegraphics[width = 9cm, height = 6.22cm]{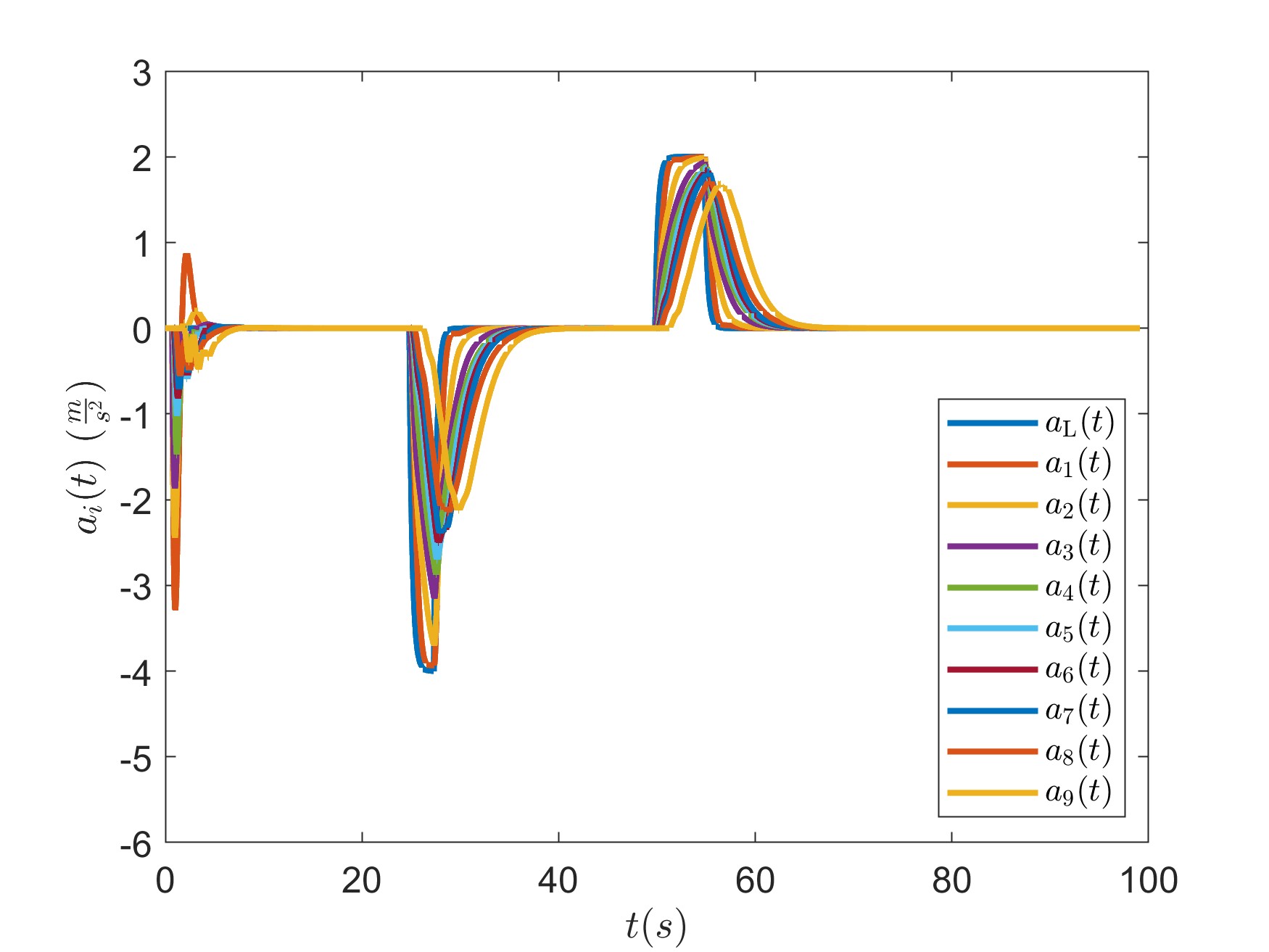}
		\includegraphics[width = 9cm, height = 6.22cm]{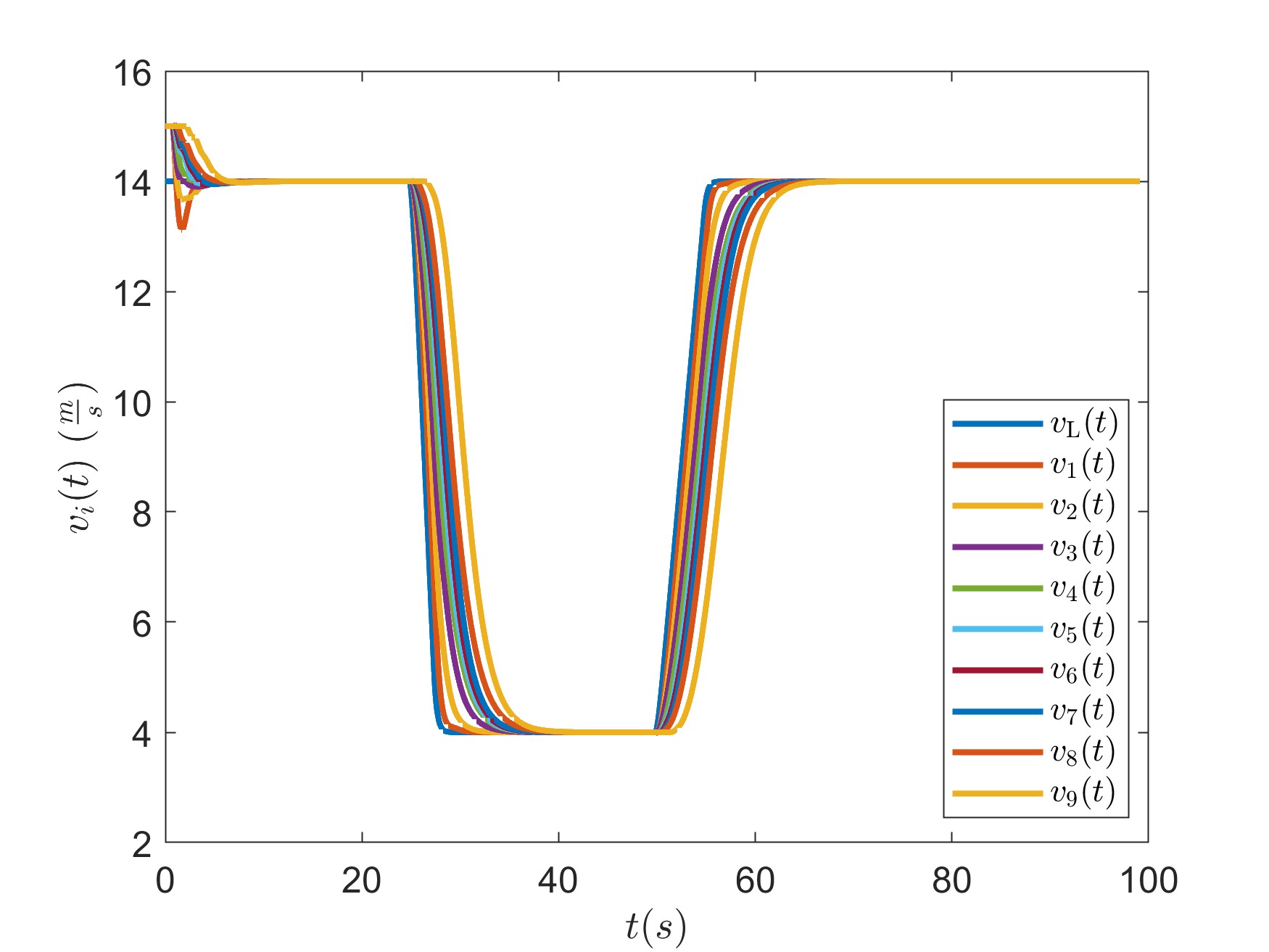}
		\includegraphics[width = 9cm, height = 6.22cm]{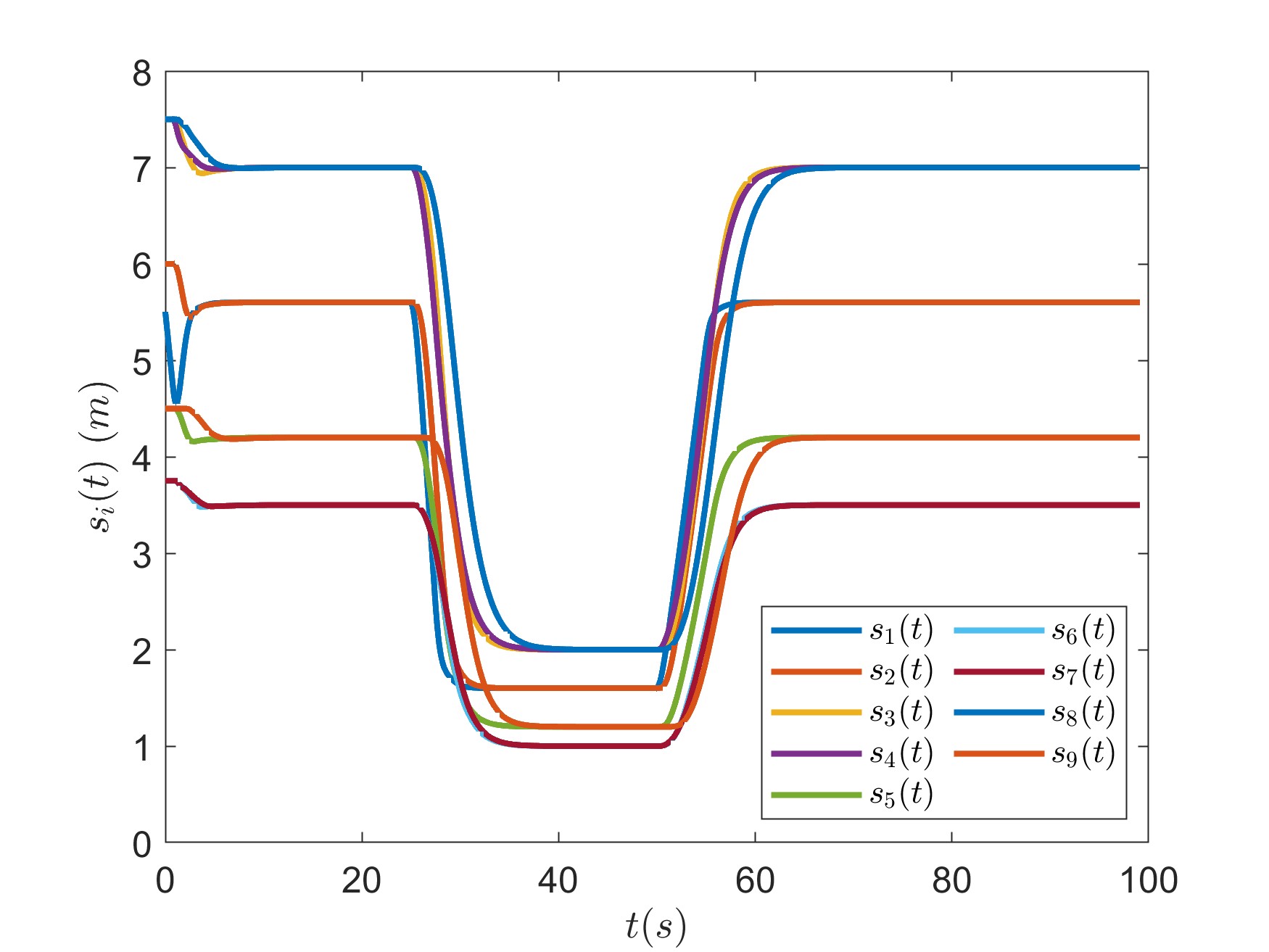}
		\caption{Acceleration (top), speed (middle), and spacing (bottom) of nine vehicles following a leader, with dynamics described by (\ref{dy1})–(\ref{dy3}), under the predictor-feedback control laws (\ref{CL})--(\ref{K1}), for parameters defined in Table~\ref{table2} and actuation delay $D = 0.7$.}\label{Fig7}
	\end{center}
\end{figure}

\begin{figure}[p]
	\begin{center}
		\includegraphics[width = 9cm, height = 6.22cm]{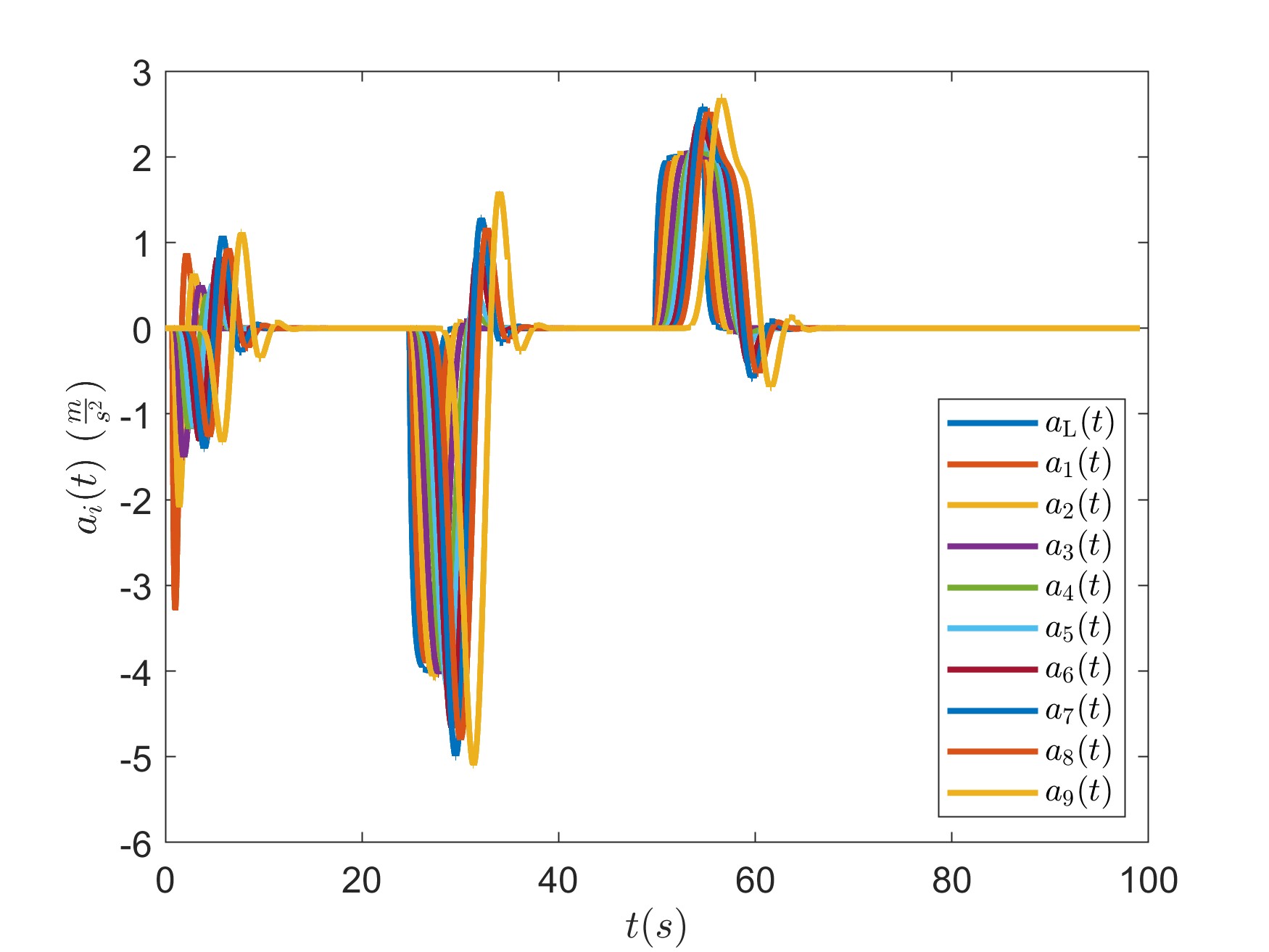}
		\includegraphics[width = 9cm, height = 6.22cm]{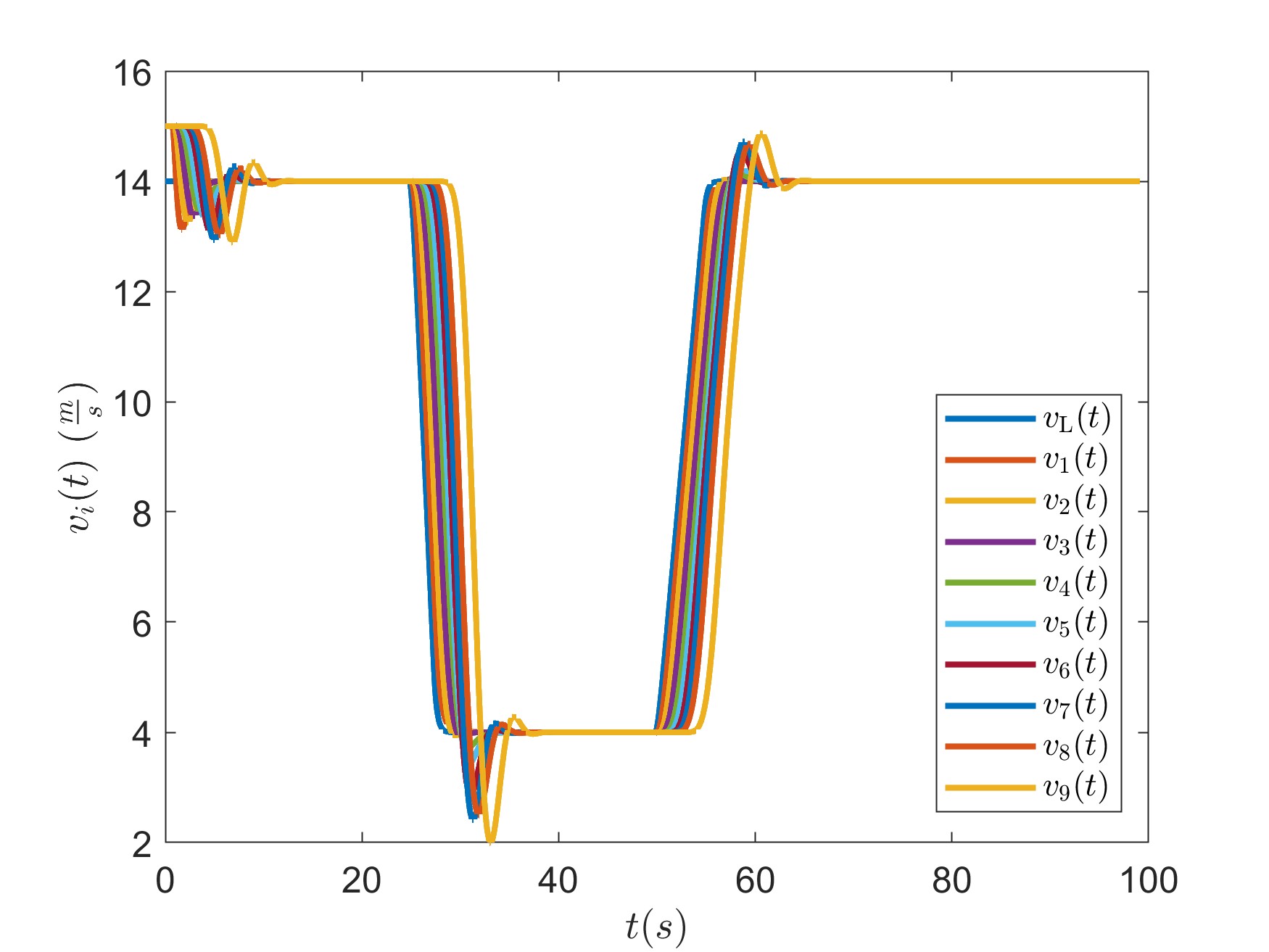}
		\includegraphics[width = 9cm, height = 6.22cm]{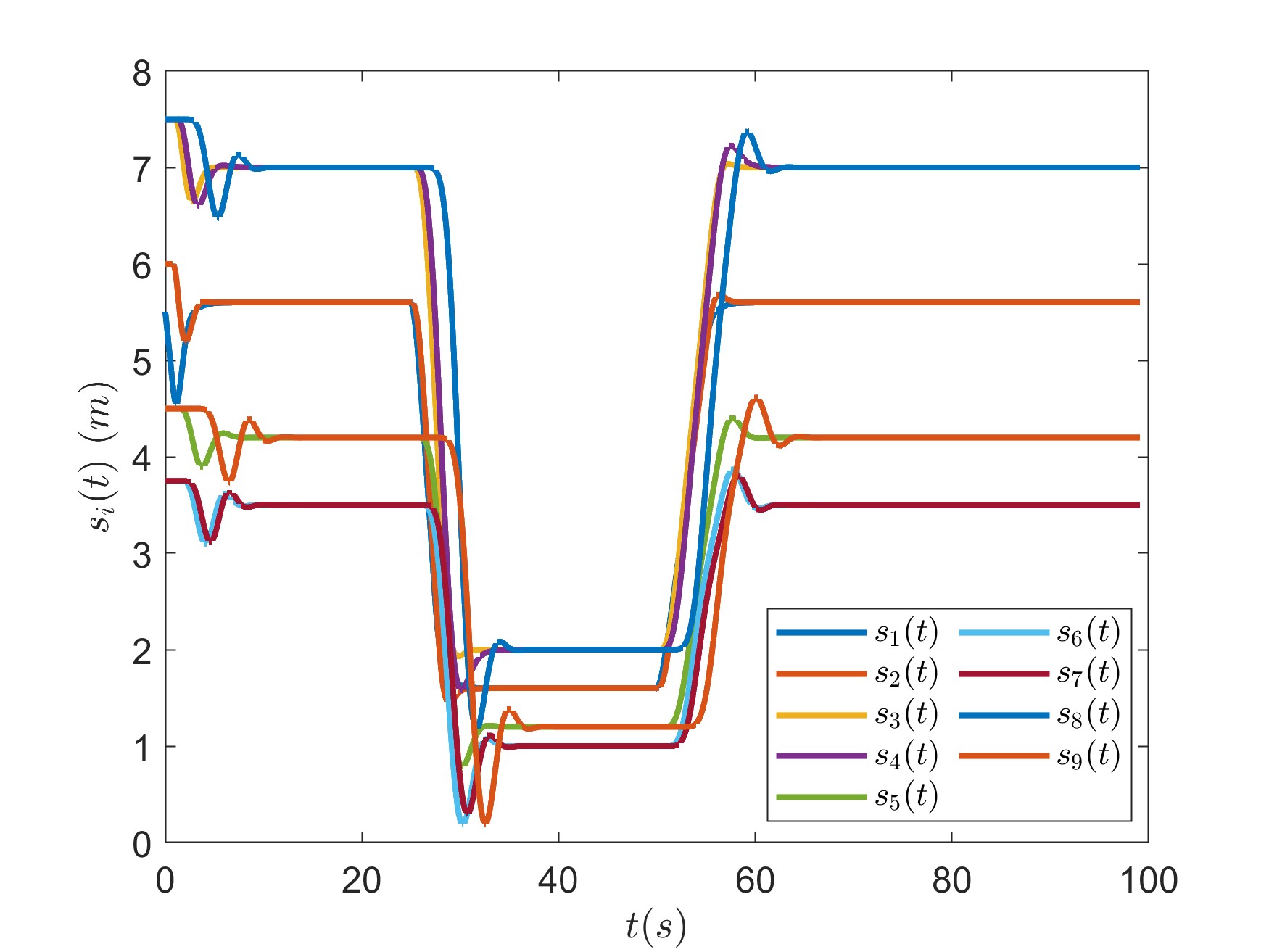}
		\caption{Acceleration (top), speed (middle), and spacing (bottom) of nine vehicles following a leader, with dynamics described by (\ref{dy1})–(\ref{dy3}), under the predictor-feedback control laws (\ref{CL})--(\ref{K1}), for parameters defined in Table~\ref{table2}, actuation delay $D = 0.7$, and for $m_i=1$, for all $i$.}\label{Fig7m}
	\end{center}
\end{figure}

    We consider a heterogeneous platoon of ten vehicles with actuation delay $D=0.7$. We choose control gains $\alpha_i=5$, $b_i=10$, and $c_i=2$ for all $i$. Initial conditions are $v_{i_0} = 15 \left(\frac{m}{s} \right)$, $v_{{\rm 0}_0} = 14 \left(\frac{m}{s} \right) $; $s_{i_0} = h_{i}v_{i_0} \nobreakspace (m)$, $s_{1_0} = 6 \nobreakspace (m)$; $a_{i_0}=0$, and $u_{i_0} \equiv 0$, for all $i$. In the present scenario we consider a case in which $\tau_i$, $h_i$, $D_{\rm c,i}$, and $m_i$ are set according to Table~\ref{table2}. The respective responses are shown in Fig.~\ref{Fig7}. We observe that there is no overshoot in the responses, due to deceleration or acceleration maneuvers performed by the leader, because the responses are string stable. Next, Fig.~\ref{Fig7m} illustrates the case in which each ego vehicle communicates with only one vehicle ahead, i.e., $m_i = 1$ for all $i$, using the same values for the rest of the parameters as in Fig.~\ref{Fig7}. The resulting controller reduces, in fact, to the one from \cite{bekiaris_cth_2023}. Although individual vehicle stability is maintained, string stability is lost, as evidenced by the overshoot observed in the velocity responses. 
    
\begin{figure}[p]
	\begin{center}
		\includegraphics[width = 9cm, height = 6.22cm]{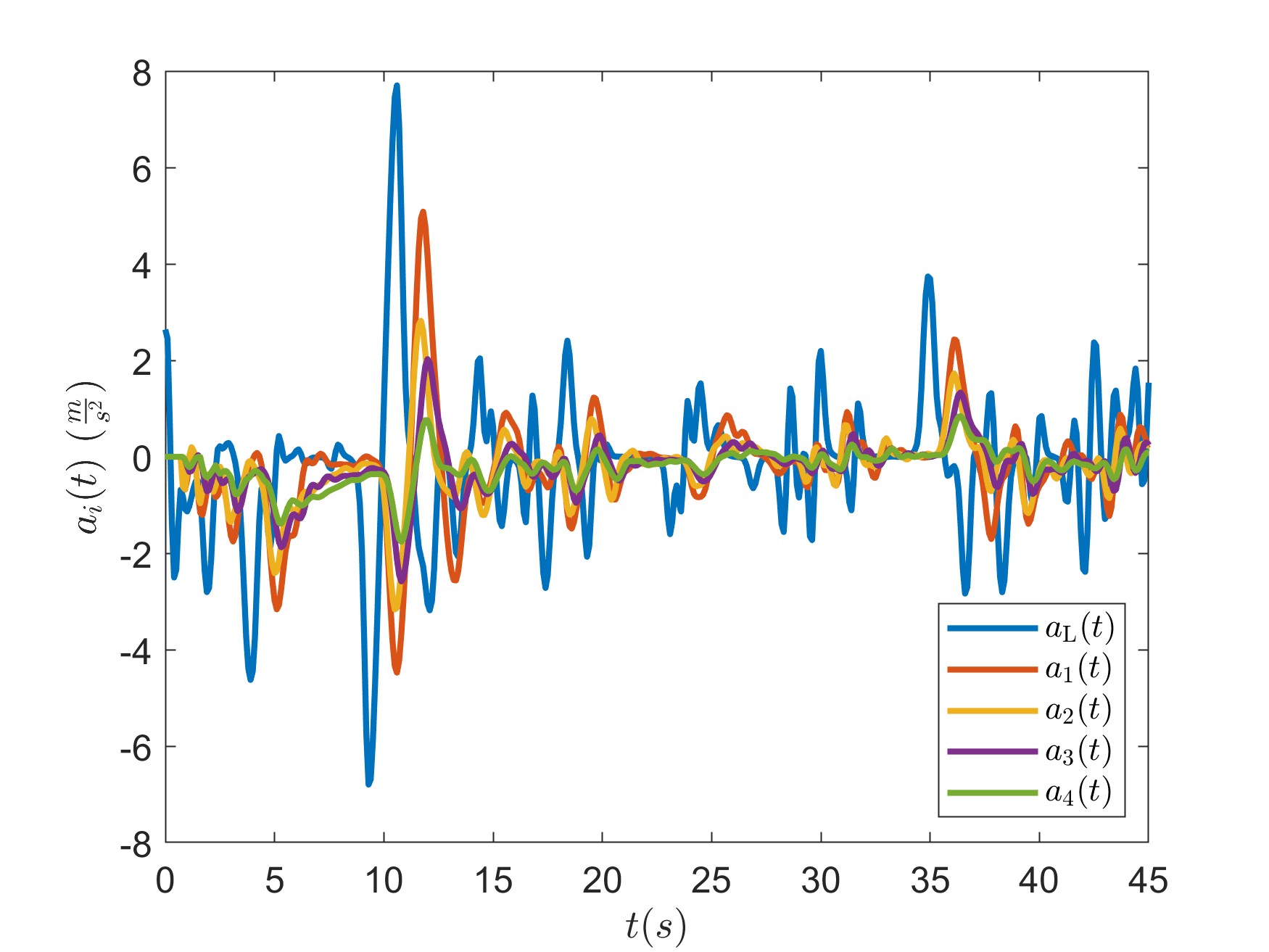}
		\includegraphics[width = 9cm, height = 6.22cm]{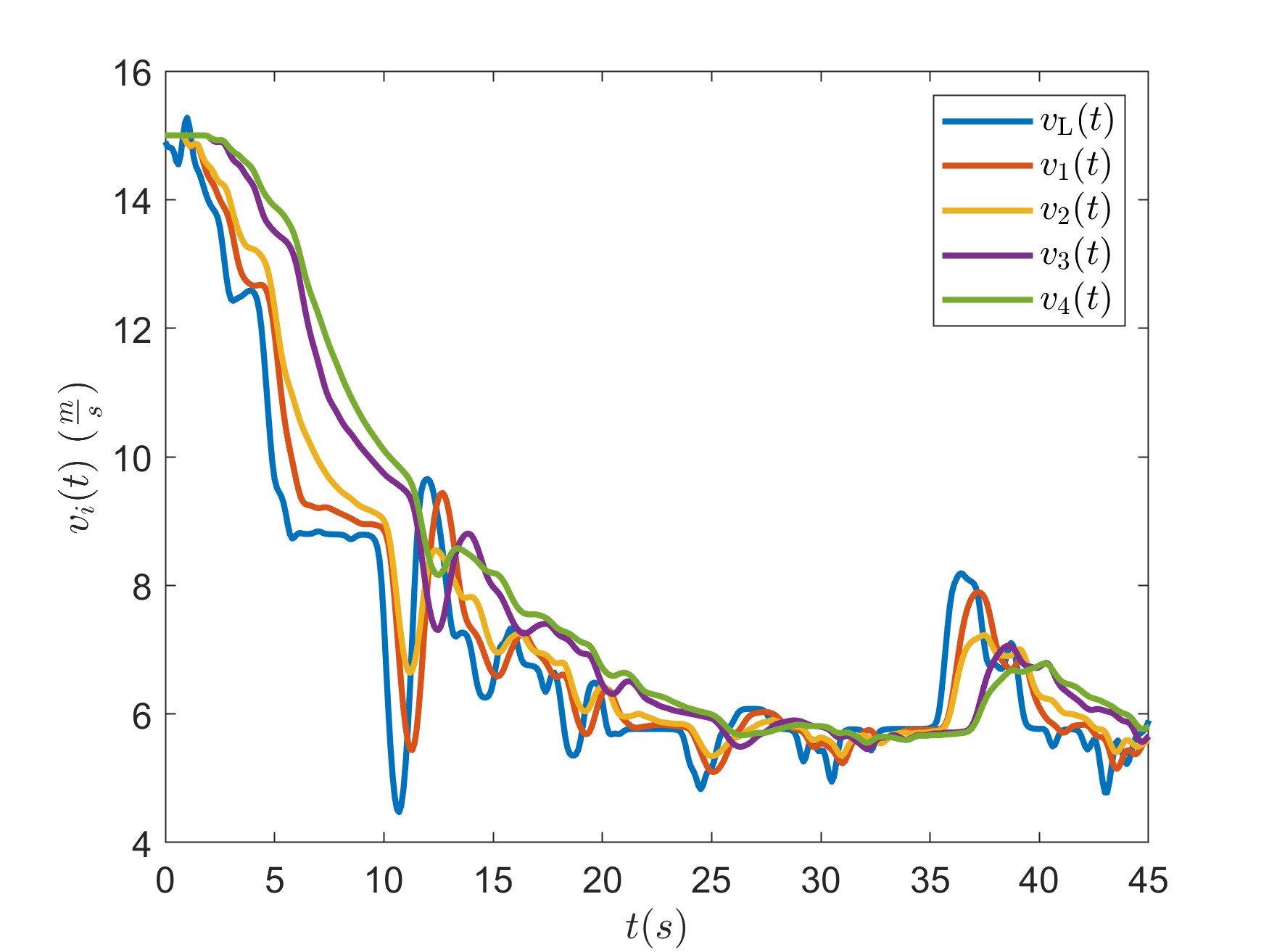}
		\includegraphics[width = 9cm, height = 6.22cm]{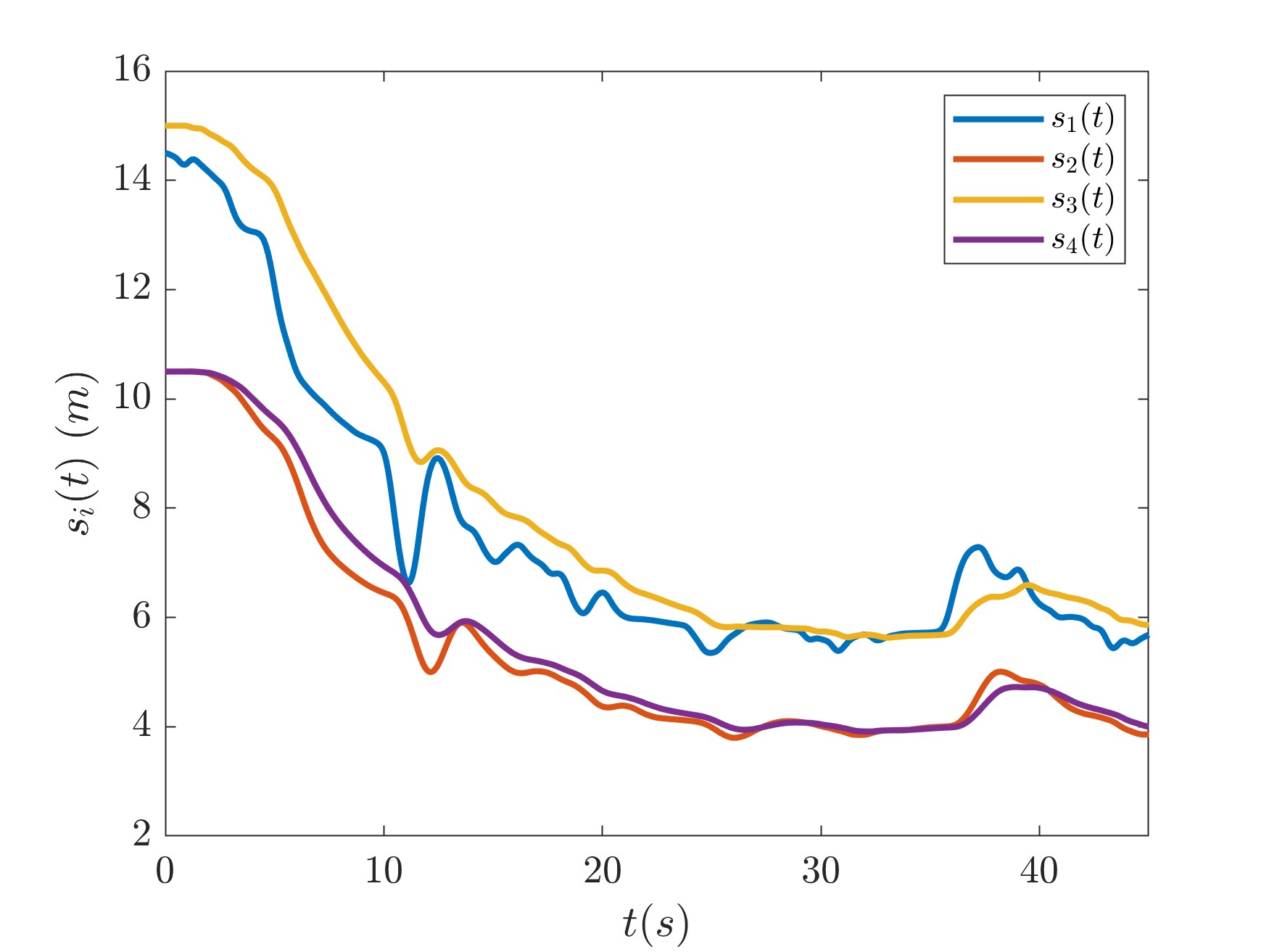}
		\caption{Acceleration (top), speed (middle), and spacing (bottom) of four vehicles following a leader whose trajectory is obtained from the trajectory of vehicle no. 1601 in the NGSIM data, under the predictor-feedback control laws (5)--(13), for parameters defined in Table~\ref{table3} and actuation delay $D = 0.7$. }\label{Fig8}
	\end{center}
\end{figure}

    Fig.~\ref{Fig8} shows the outcome of implementing the predictor–feedback CACC scheme (5)--(13) using data from the NGSIM dataset for heavily congested traffic. To evaluate the controller under realistic driving conditions, we use reconstructed trajectory data from \cite{NGSIM}, where the lead vehicle follows the recorded trajectory of vehicle No. 1601. The exact details of the simulation setup are described in \cite{predictor_feedback3}, Section~V.C. In the present scenario, we consider the case where $D = 0.7$, while the remaining parameters are set according to Table~\ref{table3}. We set $a_{i_0}(s)=0$ and $u_{i_0}(s) \equiv 0$, for vehicles $i=1, 2 , 3, 4$. While we also set $v_{i_0} = 15 \left(\frac{m}{s} \right)$, $i = 1, 2, 3, 4$ and $v_{{\rm L}_0} = 14.9 \left(\frac{m}{s} \right)$ (to match the initial speed of vehicle 1601 from NGSIM data); $s_{i_0} = h_{i}v_{i_0} = h_{i} \times 15\nobreakspace m$, $i = 2, 3, 4$, $s_{1_0} = 14.5 \nobreakspace m$. Fig.~\ref{Fig8} illustrates that the performance of the predictor-feedback CACC law with communication delay is preserved even in more realistic traffic scenarios.

\begin{table}[!ht]
	\centering
	\caption{The parameters employed for the simulation results in Fig. \ref{Fig8}.}
	\scalebox{1}{\begin{tabular}{||c| c c c c||} 
			\hline
			\backslashbox{Vehicle No.}{Parameters} & $\tau_{i}$ & $h_i$ & $D_{\rm c,i}$ & $m_i$\\ 
			\hline\hline
			0&$-$ & $-$ & $0.1$ s & $1$\\ 
			\hline
			1&$0.3$ s & $1$ s & $0.2$ s & $2$\\ 
			\hline
			2&$0.25$ s & $0.7$ s & $0.1$ s & $2$\\
			\hline
			3&$0.25$ s & $1$ s & $0.1$ s & $2$\\
			\hline
            4&$0.2$ s & $0.7$ s & $-$ & $2$ \\
			\hline
	\end{tabular}}
	\label{table3}
\end{table}

\section{Conclusions and Future Work}
In this paper, we developed a predictor-feedback CACC law relying on an MPF topology-based nominal CACC design, which achieves compensation of long actuation delay, under simultaneous presence of communication delays, and higher traffic throughput, for heterogeneous vehicular platoons. We derived analytical conditions on control/model parameters that guarantee $\mathcal{L}_2$ string stability, which we also verified numerically. Moreover, we presented consistent simulation results for a ten-vehicle platoon, including a comparison with a predictor-feedback CACC law that utilizes information only from a preceding vehicle ahead. We also validated the performance of the design developed in simulation, using real traffic data to describe the trajectory of the leading vehicle. As future work, we aim to extend our framework to account for heterogeneous (distinct) actuation delays, in each vehicle. This requires development of a completely different and more complex approach, along the lines of \cite{predictor_feedback4}.

\section*{Appendix A}
	\numberwithin{equation}{section}
	\renewcommand{\theequation}{A.\arabic{equation}}
	\setcounter{equation}{0}
	In order to study stability and string stability of speed errors propagation, we first compute the transfer functions
\begin{align}
	V_i(s) &=\sum_{n_i=1}^{m_i} G_{i,i-n_i}(s) V_{i-n_i}(s),\quad i=1,...,N, \label{VS}
\end{align}
viewing the speed of $m_i$ vehicles ahead, with which the ego vehicle $i$ communicates, as the input and the ego vehicle’s speed as the output. Taking Laplace transform of the predictor states (\ref{PF_m}) we get
\begin{align}
	Q_i(s)=& \nobreakspace e^{\Gamma_iD}\bar{X}_{i}(s)+M_{i}(s)U_{i}(s)\nonumber\\
    &+\sum_{n_i=1}^{m_i}M_{i-n_i}(s)U_{i-n_i}(s),\label{q_1s_c}
\end{align}
where for $n_i=1,\ldots,m_i$
\begin{align}
    M_{i}(s)&=(sI_{3m_i+2}-\Gamma_i)^{-1}\left( I_{3m_i+2}-e^{\Gamma_iD}e^{-sD}\right)B_{i}.\label{M_111}\\
	M_{i-n_i}(s)&=(sI_{3m_i+2}-\Gamma_i)^{-1}\left( I_{3m_i+2}-e^{\Gamma_iD}e^{-sD}\right)\nonumber\\
    &\times e^{-sD_{{\rm c},i-n_i}}B_{i-n_i}.\label{M_111_c2}
\end{align}
Then by using (\ref{q_1s_c})--(\ref{M_111_c2}) and (\ref{CL}) we get
\begin{align}
	U_i(s)=&K_i^{\rm T}\left(e^{\Gamma_iD}\bar{X}_{i}(s)+\sum_{n_i=0}^{m_i}M_{i-n_i}(s)U_{i-n_i}(s)\right).\label{u1s}
\end{align}
Using the $i$-th vehicle’s model (\ref{dy1})--(\ref{dy3}) we obtain
\begin{equation}
	\begin{bmatrix} 
		S_i(s)\\
		V_i(s)\\
		A_i(s)\\
	\end{bmatrix}=
	\begin{bmatrix} 
		-\frac{1}{s^2(s\tau_i+1)}\\
		\frac{1}{s(s\tau_i+1)}\\
		\frac{1}{s\tau_i+1}\\
	\end{bmatrix}e^{-sD}U_i(s)+\begin{bmatrix} 
		\frac{1}{s}\\
		0\\
		0\\
	\end{bmatrix}V_{i-1}(s).\label{sss}
\end{equation}
We next, using (\ref{sss}), express $\bar{X}_i(s)$ defined in (7) and involved in (\ref{u1s}), as
\begin{align}
	\bar{X}_i(s)&=P_{i}(s)e^{-sD}U_i(s)+P_{i-1,1}(s)e^{-sD}U_{i-1}(s)\nonumber\\
    &+P_{i-1,2}(s)e^{-sD_{{\rm c},i-1}}e^{-sD}U_{i-1}(s)\nonumber\\
    &+P_{i-2}(s)e^{-sD_{{\rm c},i-2}}e^{-sD}U_{i-2}(s)+\dots\nonumber\\
    &+P_{i-m_i}(s)e^{-sD_{{\rm c},i-m_i}}e^{-sD}U_{i-m_i}(s),\label{B.3}
\end{align}
where
\begin{align}
	P_{i}(s)&=\begin{bmatrix} 
		-\frac{1}{s^2(s\tau_i+1)}\\
        {\bf{0}}_{m_i-1}\\
		\frac{1}{s(s\tau_i+1)}\\
		{\bf{0}}_{m_i}\\
		\frac{1}{s\tau_i+1}\\
		{\bf{0}}_{m_i}
	\end{bmatrix},\\
	P_{i-1,1}(s)&=\begin{bmatrix} 
        \frac{1}{s^2(s\tau_{i-1}+1)}\left(1-e^{-sD_{{\rm c},i-1}}\right)\\
		{\bf{0}}_{3m_i+1}
	\end{bmatrix},\\
    P_{i-1,2}(s)&=\begin{bmatrix} 
        \frac{1}{s^2(s\tau_{i-1}+1)}\\
		-\frac{1}{s^2(s\tau_{i-1}+1)}\\
		{\bf{0}}_{m_i-2}\\
        0\\
		\frac{1}{s(s\tau_{i-1}+1)}\\
		{\bf{0}}_{m_i-1}\\
        0\\
		\frac{1}{s\tau_{i-1}+1}\\
        {\bf{0}}_{m_i-1}
	\end{bmatrix},\\
    P_{i-2}(s)&=\begin{bmatrix} 
        {\bf{0}}_{2}\\
		\frac{1}{s^2(s\tau_{i-2}+1)}\\
		{\bf{0}}_{m_i-3}\\
        {\bf{0}}_{2}\\
		\frac{1}{s(s\tau_{i-2}+1)}\\
		{\bf{0}}_{m_i-2}\\
        {\bf{0}}_{2}\\
		\frac{1}{s\tau_{i-2}+1}\\
        {\bf{0}}_{m_i-2}
	\end{bmatrix},\\
    \mathord{\vdots}\nonumber\\
    P_{i-m_i}(s)&=\begin{bmatrix} 
        {\bf{0}}_{m_i-1}\\
		\frac{1}{s^2(s\tau_{i-m_i}+1)}\\
		{\bf{0}}_{m_i}\\
		\frac{1}{s(s\tau_{i-m_i}+1)}\\
		{\bf{0}}_{m_i}\\
		\frac{1}{s\tau_{i-m_i}+1}
	\end{bmatrix}.
    \label{split1}
\end{align}
Substituting (\ref{M_111}), (\ref{M_111_c2}), and (\ref{B.3}) in (\ref{u1s}), we get
\begin{align}
	U_i(s)\delta_{i}(s) &= U_{i-1}(s)\delta_{i-1,1}(s)+U_{i-1}(s)e^{-sD_{{\rm c},i-1}}\delta_{i-1,2}(s)\nonumber\\
    &+U_{i-2}(s)e^{-sD_{{\rm c},i-2}}\delta_{i-2}(s)+\dots\nonumber\\
    &+U_{i-m_i}(s)e^{-sD_{{\rm c},i-m_i}}\delta_{i-m_i}(s),\label{Us}
\end{align}
where
\begin{align}
	\delta_{i}(s)=&1-K_i^{\rm T}P_{i}(s), \\
	\delta_{i-1,1}(s)=&K_{i}^{\rm T}e^{\Gamma_iD}e^{-sD}P_{i-1,1}(s),\label{bard}\\
	\delta_{i-1,2}(s)=&K_{i}^{\rm T}P_{i-1,2}(s),\\
    \delta_{i-2}(s)=&K_i^{\rm T}P_{i-2}(s),\\
    \mathord{\vdots}\nonumber\\
    \delta_{i-m_i}(s)=&K_i^{\rm T}P_{i-m_i}(s).\label{A.17}
\end{align}
Thus, using $V_{n_i}(s)=\frac{e^{-sD}}{s(s\tau_{n_i}+1)} U_{n_i}(s)$, for all $n_i = i,\ldots,i-m_i$, by utilizing (\ref{sss}) in (\ref{Us}) gives
\begin{align}
	\delta_{i}(s)s(s\tau_i+1)V_i(s) &=\left(\delta_{i-1,1}(s)+\delta_{i-1,2}(s)e^{-sD_{{\rm c},i-1}}\right)\nonumber\\
    &\times s(s\tau_{i-1}+1)V_{i-1}(s)\nonumber\\
    &+\sum_{n_i=2}^{m_i} \delta_{i-n_i}(s)e^{-sD_{{\rm c},i-n_i}}\nonumber\\
    &\times s(s\tau_{i-n_i}+1)V_{i-n_i}(s).\label{Gs0}
\end{align}
Hence, comparing (\ref{Gs0}) and (\ref{VS}) we arrive at
\begin{align}
G_{i,i-n_i}(s)
&=\frac{c_i s^2
+ \left( b_i - (m_i-n_i)\alpha_i \frac{h_{i-n_i}}{h_i} \right)s
+ \dfrac{\alpha_i}{h_i}}{s^3
+ \dfrac{1 + m_i \tau_i c_i}{\tau_i} s^2+ m_i(\alpha_i + b_i)s
+ \dfrac{m_i\alpha_i}{h_i}}\nonumber\\
&\times e^{-sD_{{\rm c},i-n_i}},\label{Gs}
\end{align}
for $n_i=2,\ldots,m_i$. Moreover, using the exponential series expansion, $e^{\Gamma_iD}P_{i-1,1}(s)=\sum_{k=0}^{\infty}\frac{D^k}{k!}\Gamma_i^kP_{i-1,1}(s)$ and the fact that $\Gamma_iP_{i-1,1}(s)={\bf 0}_{3m_i+2}$ along with (\ref{bard}) and (\ref{Gs0}) we get
\begin{align}
	G_{i,i-1}(s)
&=\frac{ \mu_{1,i}(s)s^2
+ \mu_{2,i}(s)s
+ \mu_{3,i}(s) }{s^3
+ \dfrac{1 + m_i \tau_i c_i}{\tau_i} s^2+ m_i(\alpha_i + b_i)s
+ \dfrac{m_i\alpha_i}{h_i}},\label{Gs_1}
\end{align}
where
\begin{align}
	\mu_{1,i}(s)&=c_ie^{-sD_{{\rm c},i-1}},\\
    \mu_{2,i}(s)&=\left( b_i - (m_i-1)\alpha_i \frac{h_{i-1}}{h_i} \right)e^{-sD_{{\rm c},i-1}},\\
    \mu_{3,i} (s)&= \frac{\alpha_i}{h_i}\left( e^{-sD_{{\rm c},i-1}} + m_ie^{-sD}(1-e^{-sD_{{\rm c},i-1}})\right).
 \end{align}
Then, since $\lim_{\omega \to 0^{+}} |G_{i,i-n_i}(j\omega)| = \frac{1}{m_i}$, string stability in $\mathcal{L}_2$, as defined in (\ref{G}), is guaranteed if and only if
\begin{equation}
\|G_{i,i-n_i}\|_{\infty} \le \frac{1}{m_i}, 
\quad 1 \le n_i \le m_i.
\end{equation}
Moreover, we can rewrite (\ref{Gs}) for $n_i\in\{2,...,m_i\}$ as
 	\begin{align}
	G_{i,i-n_i}(j\omega)=&\nobreakspace\frac{f_{1,i}(\omega)+jf_{2,i,n_i}(\omega)}{f_{3,i}(\omega)+jf_{4,i}(\omega)},\\
	f_{1,i}(\omega)=&\nobreakspace\frac{\alpha_i}{h_i}-c_i\omega^2,\\
	f_{2,i,n_i}(\omega)=&\nobreakspace \omega \left( b_i - (m_i-n_i)\alpha_i \frac{h_{i-n_i}}{h_i} \right),\\
	f_{3,i}(\omega)=&\nobreakspace - \frac{1 + m_i \tau_i c_i}{\tau_i}\,\omega^2+\frac{m_i \alpha_i}{h_i},\\
    f_{4,i}(\omega)=&\nobreakspace m_i(\alpha_i+b_i)\omega-\omega^3.
	\label{A.27}
\end{align}
Therefore, the condition for string stability becomes $m_i^2(f_{1,i}(\omega)^2+ f_{2,i,n_i}(\omega)^2) \le f_{3,i}(\omega)^2 + f_{4,i}(\omega)^2$, $\omega>0$, $i=1,...,N$, and hence, after straightforward computations, we get the following condition, which has to hold for all $\omega > 0$, $n_i=2\ldots,m_i$, and $i=1,...,N$,
\begin{align}
	&\omega^6+\omega^4\beta_i+\omega^2\gamma_{i,n_i}\ge0,
	\label{A.30}
\end{align}	
where $\beta_i$ and $\gamma_{i,n_i}$ are defined in (\ref{c_1}) and (\ref{c_2}), respectively. Using $z=\omega^2$ in relation (\ref{A.30}), we obtain
\begin{align}
	&z^2+z\beta_i+\gamma_{i,n_i}\ge0.
	\label{A.33}
\end{align}	
Relation (\ref{A.33}) (that is a second-order polynomial in $z>0$) holds for all $\omega>0$, under conditions (\ref{c_5}) or (\ref{c_3}) from Theorem~1. Individual vehicle stability follows employing, e.g., the Ruth-Hurwitz criterion in the denominator of (\ref{Gs}). Furthermore, we rewrite (\ref{Gs_1}) as
	\begin{align}
	G_{i,i-1}(j\omega)&=\nobreakspace\frac{\bar{f}_{1,i}(\omega)+j\bar{f}_{2,i}(\omega)}{f_{3,i}(\omega)+jf_{4,i}(\omega)},\\
    \bar{f}_{1,i}(\omega)
    &=\left(-\omega^2 c_i +\frac{\alpha_i}{h_i}\right) \cos(\omega D_{c,i-1})\nonumber\\
    &+ \omega\left( b_i - (m_i-1)\alpha_i \frac{h_{i-1}}{h_i}\right)\sin(\omega D_{c,i-1}) \nonumber\\
     &+ \frac{m_i\alpha_i}{h_i}\left(\cos(\omega D)
- \cos\big(\omega(D+D_{c,i-1})\big)\right),\label{B.26}\\
    \bar{f}_{2,i}(\omega)
    &=\left(\omega^2 c_i - \frac{\alpha_i}{h_i} \right) \sin(\omega D_{c,i-1})  \nonumber\\
    &+ \omega 
    \left( b_i - (m_i-1)\alpha_i \frac{h_{i-1}}{h_i} \right)
    \cos(\omega D_{c,i-1}) \nonumber\\
    & - \frac{m_i\alpha_i}{h_i}\left(\sin(\omega D)
- \sin\big(\omega(D+D_{c,i-1})\big)\right).
	\label{B.27}
\end{align}
For a given $\omega$, based on the mean-value theorem, for each $i$, there exist $\xi_i(\omega)$ and $\zeta_i(\omega)$ such that
\begin{align}
\cos(\omega D) - \cos(\omega (D+D_{{\rm c},i-1})) =&\nobreakspace \omega D_{{\rm c},i-1}\sin(\omega \xi_i(\omega)),\nonumber\\
&\xi_i \in(D,D+D_{{\rm c},i-1}),\\
\sin(\omega D) - \sin(\omega (D+D_{{\rm c},i-1})) =&\nobreakspace -\omega D_{{\rm c},i-1}\cos(\omega \zeta_i(\omega)),\nonumber\\
&\zeta_i \in(D,D+D_{{\rm c},i-1}).		
\label{B.32}
\end{align}
Hence, (\ref{B.26}), (\ref{B.27}) can also be written as
\begin{align}
\bar{f}_{1,i}(\omega)
    &=\left(-\omega^2 c_i +\frac{\alpha_i}{h_i}\right) \cos(\omega D_{c,i-1})\nonumber\\
    &+ \omega\left( b_i - (m_i-1)\alpha_i \frac{h_{i-1}}{h_i}\right)\sin(\omega D_{c,i-1}) \nonumber\\
     &+ \frac{m_i\alpha_i}{h_i}\omega D_{{\rm c},i-1}\sin(\omega \xi_i(\omega)),\\
    \bar{f}_{2,i}(\omega)
    &=\left(\omega^2 c_i - \frac{\alpha_i}{h_i} \right) \sin(\omega D_{c,i-1})  \nonumber\\
    &+ \omega 
    \left( b_i - (m_i-1)\alpha_i \frac{h_{i-1}}{h_i} \right)
    \cos(\omega D_{c,i-1}) \nonumber\\
    & + \frac{m_i\alpha_i}{h_i}\omega D_{{\rm c},i-1}\cos(\omega \zeta_i(\omega)).
\end{align}
Therefore, the condition for string stability becomes $m_i^2(\bar{f}_{1,i}(\omega)^2+ \bar{f}_{2,i}(\omega)^2) < f_{3,i}(\omega)^2 + f_{4,i}(\omega)^2$, $\omega>0$, $i=1,\dots,N$, and hence, we get the following condition that has to hold for all $\omega > 0$ 

\begin{align}
&\omega^6+\omega^4f_{5,i}+\omega^3f_{6,i}(\omega)+\omega^2f_{7,i}(\omega)+\omega f_{8,i}(\omega)>0,\label{B.21}
\end{align}	
where
\begin{align} 
f_{5,i}&=\nobreakspace \left(\frac{1+m_i\tau_i c_i}{\tau_i}\right)^2-2m_i(\alpha_i+b_i)-m_i^2 c_i^2,\\
f_{6,i}(\omega)&=\nobreakspace -m_i^2 2 c_i \frac{m_i \alpha_i}{h_i} D_{{\rm c},i-1}\left(\sin\left(\omega D_{c,i-1}\right)\cos\left(\omega \zeta_i(\omega)\right) \right. \nonumber\\
&\left.-\cos\left(\omega D_{c,i-1}\right)\sin\left(\omega \xi_i(\omega)\right)\right), \label{B.23}\\
f_{7,i}(\omega)&=\nobreakspace m_i^2(\alpha_i+b_i)^2
-2\left(\frac{1+m_i\tau_i c_i}{\tau_i}\right)\frac{m_i\alpha_i}{h_i}\nonumber\\
&-m_i^2\Bigg(-2 c_i \frac{\alpha_i}{h_i}
+ \left( b_i - (m_i-1)\alpha_i \frac{h_{i-1}}{h_i} \right)^2\nonumber\\
&+ \left( \frac{m_i \alpha_i}{h_i} D_{{\rm c},i-1} \right)^2\left(\sin^2\left(\omega \xi_i(\omega)\right)+\cos^2\left(\omega \zeta_i(\omega)\right)\right)\nonumber\\
&+2 \left( b_i - (m_i-1)\alpha_i \frac{h_{i-1}}{h_i} \right)\left( \frac{m_i \alpha_i}{h_i} D_{{\rm c},i-1} \right)\nonumber\\
&\times\left(\sin(\omega D_{c,i-1})\sin\left(\omega \xi_i(\omega)\right) \right. \nonumber \\ 
&\left.+\cos(\omega D_{c,i-1})\cos\left(\omega \zeta_i(\omega)\right)\right)\Bigg),\label{B.24}\\
f_{8,i}(\omega)&=\nobreakspace -2 m_i^2\frac{\alpha_i}{h_i}
\left( \frac{m_i \alpha_i}{h_i} D_{{\rm c},i-1} \right)\nonumber\\
&\times\left(\sin\left(\omega D_{c,i-1}\right)\cos\left(\omega \zeta_i(\omega)\right)\right. \nonumber\\ 
&\left.-\cos\left(\omega D_{c,i-1}\right)\sin\left(\omega \xi_i(\omega)\right)\right)\label{B.25}.
\end{align}
Using the facts that $|\sin(x)| \leq |x|$, for all $x \in \mathbb{R} $, that $\omega,\xi_i>0$, and that $\xi_i < D+D_{{\rm c},i-1}$ we get from (\ref{B.23}), (\ref{B.25})
\begin{align}
	f_{6,i}(\omega)&\geq -\omega m_i^2 2 c_i \frac{m_i \alpha_i}{h_i} D_{{\rm c},i-1}(2D_{{\rm c},i-1}+D),\\
	f_{8,i}(\omega)&\geq -2\omega m_i^2\frac{\alpha_i}{h_i}
\left( \frac{m_i \alpha_i}{h_i} D_{{\rm c},i-1} \right)(2D_{{\rm c},i-1}+D).
	\label{B.41}
\end{align}
Thus, condition (\ref{B.21}) is satisfied if for all $\omega>0$
\begin{align}
	&\omega^4+\omega^2\left( f_{5,i}-m_i^2 2 c_i \frac{m_i \alpha_i}{h_i} D_{{\rm c},i-1} (2D_{{\rm c},i-1}+D)\right)\nonumber\\
	&+ f_{7,i}(\omega)-2m_i^2\frac{\alpha_i}{h_i}
\left( \frac{m_i \alpha_i}{h_i} D_{{\rm c},i-1} \right)(2D_{{\rm c},i-1}+D)>0.
	\label{B.43}
\end{align}	
Since from (\ref{B.24}) we have that
\begin{align}
	f_{7,i}(\omega)&\geq\, m_i^2(\alpha_i+b_i)^2
-2\left(\frac{1+m_i\tau_i c_i}{\tau_i}\right)
\frac{m_i\alpha_i}{h_i}\nonumber\\
&-m_i^2\Bigg(
-2 c_i \frac{\alpha_i}{h_i}
+ \left( b_i - (m_i-1)\alpha_i \frac{h_{i-1}}{h_i} \right)^2
\nonumber\\
&+ 2\left( \frac{m_i \alpha_i}{h_i} D_{{\rm c},i-1} \right)^2
+ 4 \left| b_i - (m_i-1)\alpha_i \frac{h_{i-1}}{h_i} \right|\nonumber\\
&\times\left( \frac{m_i \alpha_i}{h_i} D_{{\rm c},i-1} \right)\Bigg),
	\label{B.44}
\end{align}
the condition for string stability becomes 
\begin{align}
	&z^2+z\bar{\beta}_i+\bar{\gamma}_{i}\ge0,
\end{align}	
where $\bar{\beta}_i$ and $\bar{\gamma}_{i}$ are defined in (\ref{c_0}) and (\ref{c_02}), respectively, which is satisfied under (\ref{c_5}) or (\ref{c_3}).

We next establish regulation. Using (\ref{VS}), we prove (since $G_{i,i-n_i}(0)=\frac{1}{m_i}$, for $1\leq n_i\leq m_i$) that 
\begin{align}
\lim_{t\to+\infty} v_i(t) = \lim_{t\to+\infty} v_{i-1}(t)=v_{\rm ss},
\end{align}	
for a constant leader's speed $v_{\rm ss}$. Indeed, letting $V_0=\frac{v_{\rm ss}}{s}$, then by (\ref{VS}) we have 
\begin{align}
V_1(s)=G_{1,0}(s)V_0(s),
\end{align}	
and thus, since from (\ref{Gs_1}) we get $G_{1,0}(0)=1$ (as $m_1=1$), it follows that\footnote{From (\ref{Gs}), all poles of $sV_1$ are on the left half-plane.}
\begin{align}
\lim_{s\to0}s V_1(s)=\lim_{s\to0}s V_0(s). \label{V}
\end{align}	
Moreover, using (\ref{VS}) we have when $m_2=~2$\footnote{We note that the exact same analysis can be carried out in the simpler case $m_i<i$.}
\begin{align}
V_2(s)=G_{2,1}(s)V_1(s)+G_{2,0}(s)V_0(s).
\end{align}	
Since from (\ref{Gs}) we get $G_{2,1}(0)=\frac{1}{2}$, $G_{2,0}(0)=\frac{1}{2}$ (as $m_2=2$), using (\ref{V}) it follows that
\begin{align}
\lim_{s\to0}s V_2(s)=\lim_{s\to0}\frac{s}{2} V_1(s)+\lim_{s\to0}\frac{s}{2} V_0(s)=v_{\rm ss}. 
\end{align}	
Proceeding in the exact same manner using (\ref{VS}) and (\ref{Gs}), we conclude that
\begin{align}
\lim_{t\to \infty}v_i(t)= v_{\rm ss},\quad i=1,\dots,N. 
\end{align}	
Moreover, with (\ref{dy1}) and (\ref{VS}), we have
    \begin{align}
    \lim_{s\to 0}sS_1(s)&=\lim_{s\to0}s\left(\frac{V_0(s)-V_1(s)}{s}\right)\nonumber\\
    &=\lim_{s\to0}v_{ss}\left(\frac{1-G_{1,0}(s)}{s}\right).
\end{align}
Thus, as $G_{1,0}(0)=1$\footnote{\label{note1}In fact, the transfer functions  $\frac{1-G_{1,0}(s)}{s}$ and $\frac{G_{1,0}(s)-G_{2,1}(s)G_{1,0}(s)-G_{2,0}(s)}{s}$ feature a zero-pole cancellation at $s=0$, which implies that all their poles lie in the left half-plane.} we have
\begin{equation}
	\lim_{s\to0} sS_1(s)=-\lim_{s\to0} v_{\rm ss}G_{1,0}^{'}(s).
\end{equation}
Since $\lim_{s\to0} G_{1,0}^{'}(s)= -h_1$ and $\lim_{s\to0} sV_1(s)=v_{\rm ss}$, the steady-state spacing error is calculated as
\begin{equation}
	\lim_{t\to+\infty}\left( s_1(t)-h_1v_1(t)\right) = \lim_{s\to 0}s(S_1(s)-h_1V_1(s))=0.
\end{equation}
Then for $i=2$, we have using (\ref{VS}) that
\begin{align}
    \lim_{s\to 0}sS_2(s)&=\lim_{s\to0}s\left(\frac{V_1(s)-V_2(s)}{s}\right)=\lim_{s\to0}v_{ss}\nonumber\\
    &\times\left(\frac{G_{1,0}(s)-G_{2,1}(s)G_{1,0}(s)-G_{2,0}(s)}{s}\right).
\end{align}
Thus, as $G_{1,0}(0)=1$, $G_{2,1}(0)=\frac{1}{2} $, $G_{2,0}(0)=\frac{1}{2}^{\ref{note1}}$ we have
\begin{align}
	\lim_{s\to0} sS_2(s)&=-\lim_{s\to0} v_{\rm ss}\left(G_{1,0}^{'}(s)-G_{2,1}^{'}(s)G_{1,0}(s)\right.\nonumber\\
    &\left.-G_{2,1}(s)G_{1,0}^{'}(s)-G_{2,0}^{'}(s)\right).
\end{align}
Since $\lim_{s\to0} G_{1,0}^{'}(s)-G_{2,1}^{'}(s)G_{1,0}(s)-G_{2,1}(s)G_{1,0}^{'}(s)-G_{2,0}^{'}(s)= -h_2$, the steady-state spacing error is calculated as
\begin{equation}
	\lim_{t\to+\infty}\left( s_2(t)-h_2v_2(t)\right) = \lim_{s\to 0}s(S_2(s)-h_2V_2(s))=0.
\end{equation}
Proceeding in the exact same manner for all $i=3,\dots,N$, we complete the proof, noting that individual vehicles' stability and (\ref{dy2}) imply that ${\rm lim}_{t\to+\infty}a_i(t)=0$, $i=1,\ldots,N$.

\vspace*{-\baselineskip}
\begin{IEEEbiography}[{\includegraphics[width=1in,height=1.25in,clip,keepaspectratio]{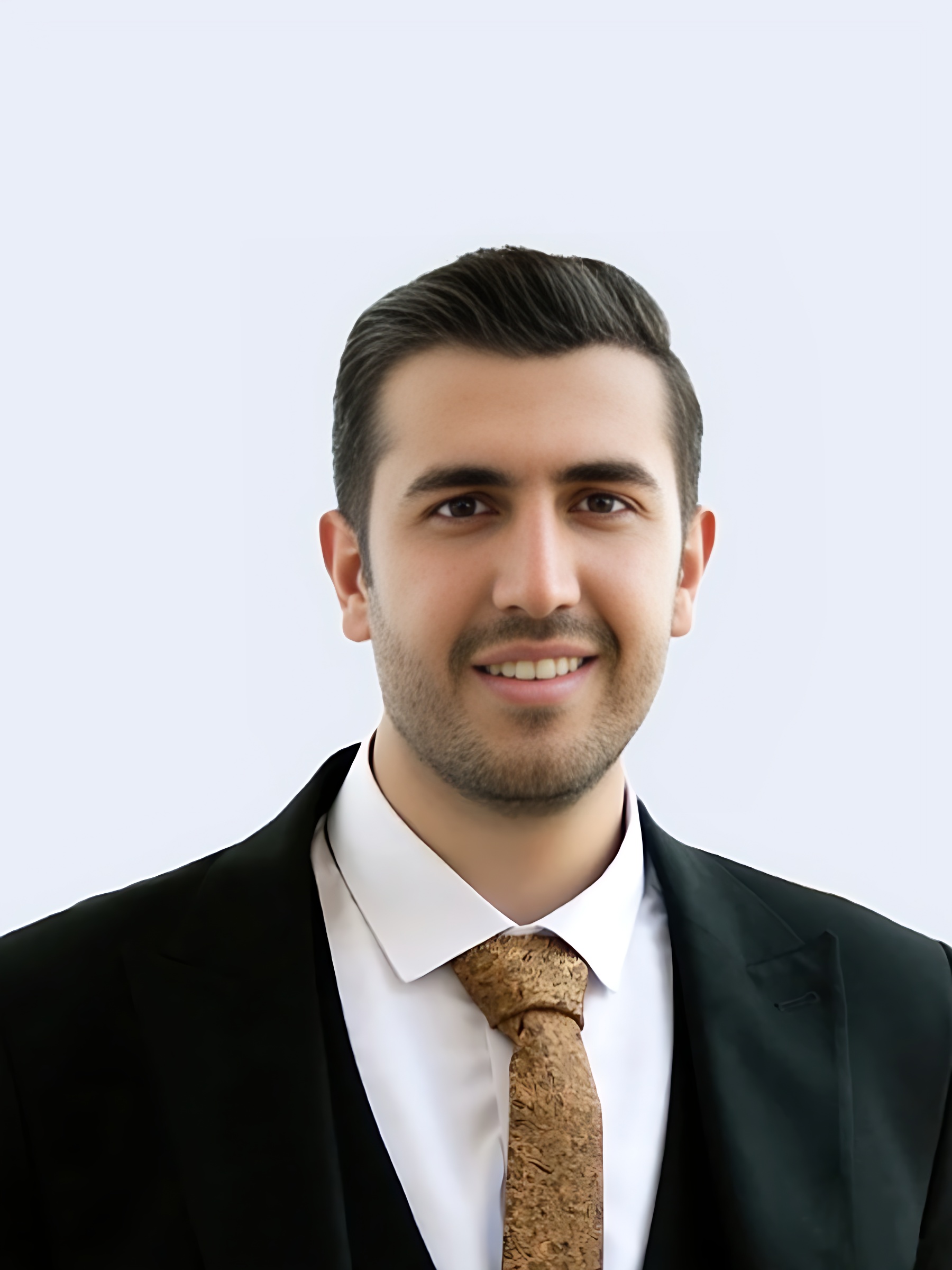}}]{Amirhossein Samii} received the B.Sc. degree in electrical and computer engineering from Isfahan University of Technology, Iran, in 2018, the M.Sc. degree in electrical and computer engineering from K. N. Toosi University of Technology, Iran, in 2021, and the Ph.D. degree in electrical and computer engineering from Technical University of Crete, Greece, in 2025, where he is currently a Post-Doctoral Researcher. His research interests include delay systems, adaptive control, and nonlinear dynamics, and their applications to connected and automated vehicles.
\vspace*{-\baselineskip}
\end{IEEEbiography}

\begin{IEEEbiography} 
[{\includegraphics[width=1in,height=1.25in,clip,keepaspectratio]{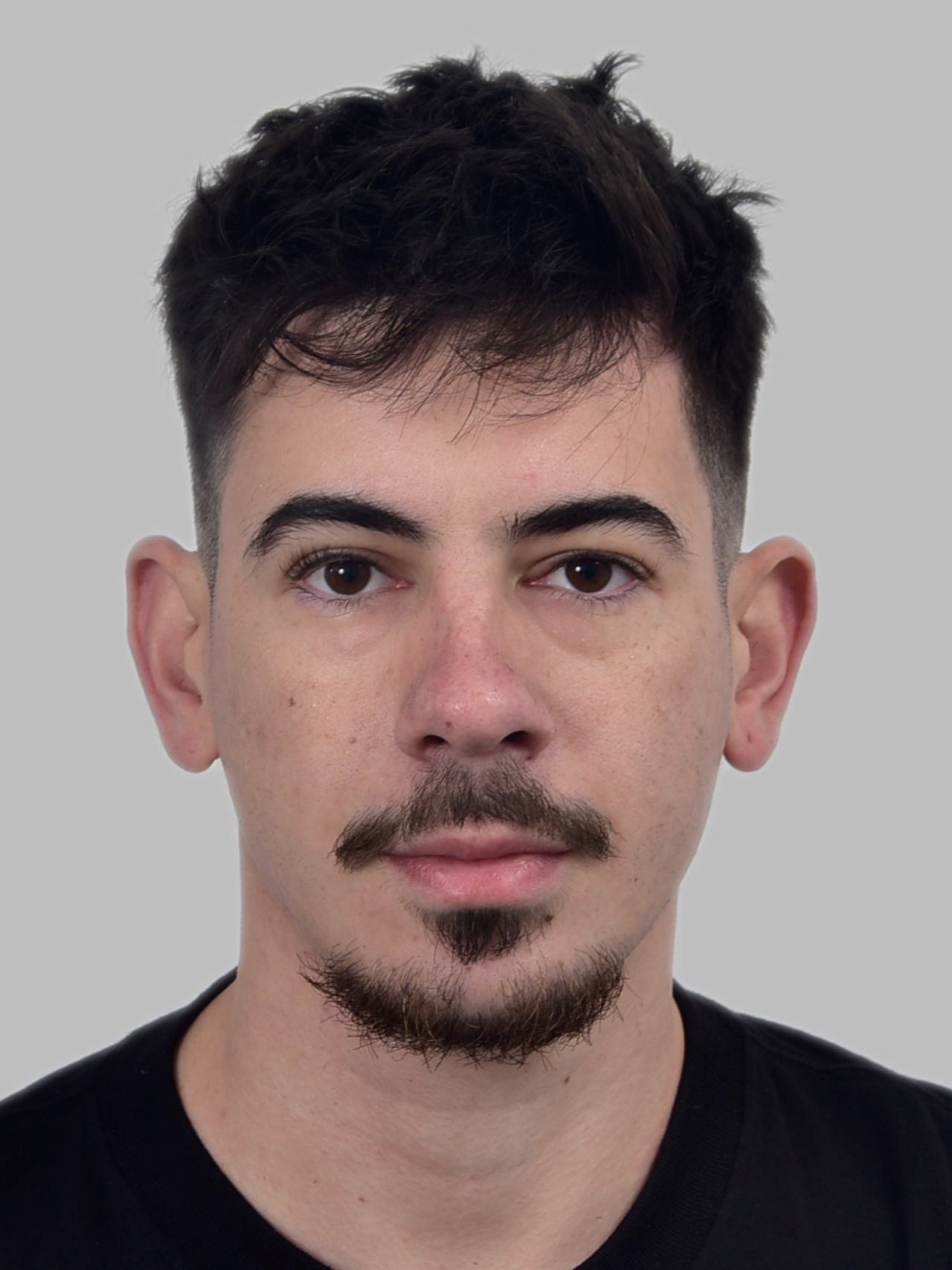}}]{Dimitrios Angelopoulos} received the diploma in electrical and computer engineering from Technical University of Crete, Greece, in 2026. His research interests include nonlinear, delay and, multi-agent systems and their applications to automated vehicles.
\vspace*{-\baselineskip}
\end{IEEEbiography}

\begin{IEEEbiography}[{\includegraphics[width=1in,height=1.25in,clip,keepaspectratio]{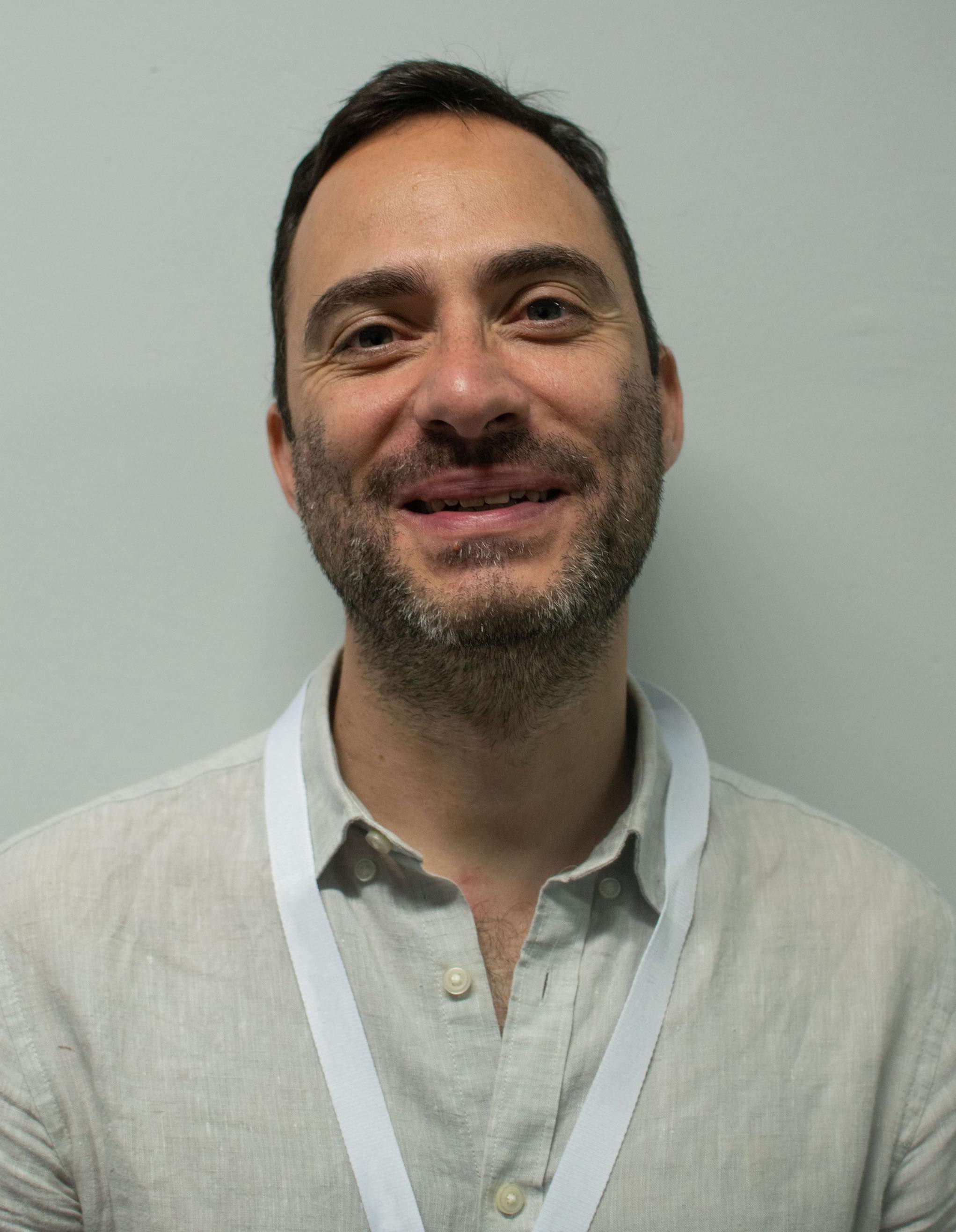}}]{Nikolaos Bekiaris-Liberis}(Senior Member, IEEE)  \\
	Nikolaos Bekiaris-Liberis received the Ph.D. degree in aerospace engineering from University of California, San Diego in 2013. From 2013 to 2014, he was a Post-Doctoral Researcher with University of California, Berkeley. From 2019 to 2022, he was an Assistant Professor, from 2017 to 2019, he was a Marie Sklodowska-Curie Fellow, and from 2014 to 2017, he was a Research Associate with the Technical University of Crete, Greece, where he is currently an Associate Professor with the Department of Electrical and Computer Engineering. His research interests include nonlinear delay, switched, and distributed parameter systems, and their applications to transport systems.
	
	Nikolaos Bekiaris-Liberis serves as Associate Editor for Automatica and IEEE Transactions on Automatic Control, as well as Senior Editor for IEEE Transactions on Intelligent Transportation Systems. He received the Chancellor’s Dissertation Medal in Engineering from University of California, San Diego in 2014 and the George N. Saridis Outstanding Research Paper Award in 2019. He was a recipient of a 2016 Marie Sklodowska-Curie Individual Fellowship Grant and he received a 2022 European Research Council (ERC) Consolidator Grant.
\end{IEEEbiography}


\begin{thebibliography}{00}

\bibitem{nominal_control_1}  
E. Abolfazli, B. Besselink, and T. Charalambous, “Minimum time headway in platooning systems under the MPF topology for different wireless communication scenarios,” {\em IEEE Transactions on Intelligent Transportation Systems}, vol. 24, pp. 4377-4390, 2023. 

\bibitem{predictor_based_0}
Z. Artstein, “Linear systems with delayed controls: A reduction", {\em IEEE Transactions on Automatic Control}, vol. 27, pp. 869-879, 1982.

\bibitem{bekiaris_predictor_2018}
N. Bekiaris-Liberis, C. Roncoli, and M. Papageorgiou, “Predictor-based adaptive cruise control design,” {\em IEEE Transactions on Intelligent Transportation Systems}, vol. 19, no. 10, pp. 3181-3195, Oct. 2018.

\bibitem{b2} 
N. Bekiaris-Liberis and M. Krstic, {\em Nonlinear Control Under Nonconstant Delays}, SIAM, 2013.

\bibitem{bekiaris_cth_2023}
N. Bekiaris-Liberis, “Robust string stability and safety of CTH predictor-feedback CACC,” {\em IEEE Transactions on Intelligent Transportation Systems}, vol. 24, no. 8, pp. 8209-8221, 2023.

\bibitem{nominal_control_2}  
Y. Bian, Y. Zheng, W. Ren, S. Eben Li, J. Wang, and K. Li, “Reducing time headway for platooning of connected vehicles via V2V communication,” {\em Transportation Research Part C: Emerging Technologies}, vol. 102, pp. 87-105, 2019.  

\bibitem{speed_error1}  
A. Bose and P. A. Ioannou, “Analysis of traffic flow with mixed manual and semiautomated vehicles,” {\em IEEE Transactions on Intelligent Transportation Systems}, vol. 4, pp. 173-188, 2003.  

\bibitem{caiazzo_dos_2023}
B. Caiazzo, D. G. Lui, A. Mungiello, A. Petrillo, and S. Santini, “On the resilience of autonomous connected vehicles platoon under DoS attacks: A predictor-based sampled data control,” in {\em IEEE Conference on Intelligent Transportation Systems}, Bilbao, Spain, 2023.

\bibitem{davis_2021}
L. C. Davis, “Method of compensation for the mechanical response of connected adaptive cruise control vehicles,” {\em Physica A: Statistical Mechanics and its Applications}, vol. 562, 2021.

\bibitem{dutch}  
R. de Haan, T. P. J. van der Sande, E. Lefeber, and I. J. M. Besselink, “Cooperative adaptive cruise control for heterogeneous platoons with delays: Controller design and experiments,” {\em IEEE Transactions on Control Systems Technology}, vol. 33, no. 4, pp. 1361-1371, 2025.  

\bibitem{dutch2}  
R. de Haan, L. Redi, T. van der Sande, and E. Lefeber, “Platooning of heterogeneous vehicles with actuation delays: Experimental results,” {\em IFAC-PapersOnLine}, vol. 58, no. 27, pp. 131-136, 2024.  

\bibitem{dehaan_observer_2023}
R. de Haan, T. van der Sande, and E. Lefeber, “Observer based cooperative adaptive cruise control for heterogeneous vehicle platoons with actuator delay,” in {\em IEEE Conference on Intelligent Transportation Systems} pp. 5204-5209, Bilbao, Spain, 2023.

\bibitem{speed_error2}  
J. I. Ge and G. Orosz, “Dynamics of connected vehicle systems with delayed acceleration feedback,” {\em Transportation Research Part C: Emerging Technologies}, vol. 46, pp. 46-64, 2014.

\bibitem{huang_ren_1998}
S. Huang and W. Ren, “Autonomous intelligent cruise control with actuator delays,” {\em Journal of Intelligent and Robotic Systems}, vol. 23, no. 1, pp. 27-43, 1998.

\bibitem{string_stability}  
S. Konduri, P. R. Pagilla, and S. Darbha, “Vehicle platooning with multiple vehicle look-ahead information,” {\em IFAC-PapersOnLine}, vol. 50, no. 1, pp. 5768-5773, 2017.

\bibitem{Kristic} 
M. Krstic, “Delay Compensation for Nonlinear, Adaptive, and PDE Systems,” {\em Birkhäuser Boston, MA}, 2009.

\bibitem{molnar_predictor_2018}
T. G. Molnar, W. B. Qin, T. Insperger, and G. Orosz, “Application of predictor feedback to compensate time delays in connected cruise control,” {\em IEEE Transactions on Intelligent Transportation Systems}, vol. 19, no. 2, pp. 545-559, 2018.

\bibitem{NGSIM} 
	M. Montanino and V. Punzo, “Trajectory data reconstruction and simulation-based validation against macroscopic traffic patterns,” {\em Transportation Research Part B: Methodological}, vol. 80, pp. 82--106, 2015.

\bibitem{molnar_safety_2023}
T. G. Molnar, A. K. Kiss, A. D. Ames, and G. Orosz, “Safety-critical control with input delay in dynamic environment,” {\em IEEE Transactions on Control Systems Technology}, vol. 31, no. 4, pp. 1507-1520, 2023.

\bibitem{oncu_string_2014}
S. Oncü, J. Ploeg, N. van de Wouw, and H. Nijmeijer, “Cooperative adaptive cruise control: Network-aware analysis of string stability,” {\em IEEE Transactions on Intelligent Transportation Systems}, vol. 15, no. 4, pp. 1527-1537, 2014.

\bibitem{Molnar Book}
G. Orosz and T. G. Molnár, “Dynamics and Control of Connected Vehicles,” {\em Springer, Cham}, 2025.

\bibitem{Pan}  
D. Pan, X. Ge, D. Ding, and X.-M. Zhang, “String-stable platooning control of connected automated vehicles under non-uniform stochastic communication delays,” {\em International Journal of Robust and Nonlinear Control}, pp. 1--18, 2026.

\bibitem{italian}  
A. Petrillo, A. Salvi, S. Santini, and A. S. Valente, “Adaptive multi-agent synchronization for collaborative driving of autonomous vehicles with multiple communication delays,” {\em Transportation Research Part C: Emerging Technologies}, vol. 86, pp. 372-392, 2018.  

\bibitem{L_stability}  
J. Ploeg, N. van de Wouw, and H. Nijmeijer, “Lp string stability of cascaded systems: Application to vehicle platooning,” {\em IEEE Transactions on Control Systems Technology}, vol. 22, pp. 786–793, 2014.  

\bibitem{robust}
A. Samii and N. Bekiaris-Liberis, “Robustness of string stability of linear predictor-feedback CACC to communication delay", {\em IEEE International Conference on Intelligent Transportation Systems}, Bilbao, Spain, pp.  5204--5209, 2023.

\bibitem{predictor_feedback3}  
A. Samii and N. Bekiaris-Liberis, “Simultaneous compensation of actuation and communication delays for heterogeneous platoons via predictor-feedback CACC with integral action,” {\em IEEE Transactions on Intelligent Vehicles}, vol. 9, pp. 5618-5630, 2024.

\bibitem{predictor_feedback4}  
A. Samii and N. Bekiaris-Liberis, “Exact predictor-feedback CACC of heterogeneous vehicular platoons with distinct actuation delays,” {\em IEEE Transactions on Intelligent Transportation Systems}, to appear, 2026.

\bibitem{predictor_feedback5}
A. Samii and N. Bekiaris-Liberis, “Predictor-based CACC design for heterogeneous vehicles with distinct input delays,” {\em IEEE Open Journal of Intelligent Transportation Systems}, vol. 5, pp. 783-796, 2024.

\bibitem{CDC}
A. Samii, R. de Haan, and N. Bekiaris-Liberis, “Experimental implementation and validation of predictor-based CACC for vehicular platoons with distinct actuation delays,” {\em IEEE Conference on Decision and Control}, Rio de Janeiro, 2025.

\bibitem{van Arem} 
B. van Arem, C. J. G. van Driel and R. Visser, “The impact of cooperative adaptive cruise control on traffic-flow characteristics”, {\em IEEE Trans. Intell. Transp. Syst.}, vol. 7, pp. 429-436, 2006.

\bibitem{wang_delay_2016}
M. Wang {\em et al.}, “Delay-compensating strategy to enhance string stability of autonomous vehicle platoons,” {\em Transportmetrica B: Transport Dynamics}, vol. 6, pp. 211-229, 2016.




\bibitem{xiao_gao_2011}
L. Xiao and F. Gao, “Practical string stability of platoon of adaptive cruise control vehicles,” {\em IEEE Transactions on Intelligent Transportation Systems}, vol. 12, no. 4, pp. 1184-1194, 2011.

\bibitem{dy1} 
H. Xing, J. Ploeg, and H. Nijmeijer, “Smith predictor compensating for vehicle actuator delays in cooperative ACC systems,” {\em IEEE Transactions on Vehicular Technology}, vol. 68, pp. 1106-1115, 2018.

\bibitem{dy2} 
H. Xing, J. Ploeg, and H. Nijmeijer, “Compensation of communication delays in a cooperative ACC system,” {\em IEEE Transactions on Vehicular Technology}, vol. 69, pp. 1177-1189, 2019.

\bibitem{yanakiev_2001}
D. Yanakiev and I. Kanellakopoulos, “Longitudinal control of automated CHVs with significant actuator delays,” {\em IEEE Transactions on Vehicular Technology}, vol. 50, no. 5, pp. 1289-1297, 2001.

\bibitem{zhang_switched_2023}
H. Zhang, J. Liu, Z. Wang, C. Huang, and H. Yan, “Adaptive switched control for connected vehicle platoon with unknown input delays,” {\em IEEE Transactions on Cybernetics}, vol. 53, no. 3, pp. 1511-1521, 2023.

\bibitem{speed_error3} 
Y. Zhang, Y. Bai, J. Hu, D. Cao, and M. Wang, “Memory-anticipation strategy to compensate for communication and actuation delays for string-stable platooning,” {\em IEEE Transactions on Intelligent Vehicles}, vol. 8, pp. 1145-1155, 2022.

\bibitem{zhao_yu_2024}
C. Zhao and H. Yu, “Robust safety for mixed-autonomy traffic with delays and disturbances,” {\em IEEE Transactions on Intelligent Transportation Systems}, vol. 25, no. 11, pp. 16522-16535, 2024.

\bibitem{V2V} 
Y. Zheng, S. Eben Li, J. Wang, D. Cao, and K. Li, “Stability and scalability of homogeneous vehicular platoon: Study on the influence of information flow topologies,” {\em Transportation Research Part C: Emerging Technologies}, vol. 17, pp. 14-26, 2016.	
\end{thebibliography}
\end{document}